\documentclass[prl,aps,twocolumn,showpacs,showkeys,nofootinbib,preprintnumbers,
superscriptaddress,amsmath,amssymb,floatfix,footinbib,a4paper]{revtex4-1}

\usepackage{graphicx}
\usepackage[usenames, dvipsnames]{color}
\usepackage{epsfig}
\usepackage{wasysym}

\newcommand{\br}{{\bf r}}
\newcommand{\bx}{{\bf x}}
\newcommand{\be}{{\bf e}}
\newcommand{\bn}{{\bf n}}

\newcommand{\bv}{{\bf v}}

\newcommand{\bu}{{\bf u}}
\newcommand{\bF}{{\bf F}}

\newcommand{\eq}[1]{Eq.~(\ref{#1})}
\newcommand{\eqs}[1]{Eqs.~(\ref{#1})}
\newcommand{\fig}[1]{Fig.~\ref{#1}}

\begin{document}

\title{Effective interaction between active colloids and fluid interfaces \\ 
induced by Marangoni flows
}

\author{Alvaro Dom\'\i nguez}
\email{\texttt{dominguez@us.es}}
\affiliation{F\'\i sica Te\'orica, Universidad de Sevilla, Apdo.~1065, 
41080 Sevilla, Spain}

\author{P. Malgaretti}
\email{\texttt{malgaretti@is.mpg.de}}

\author{M. N. Popescu}

\author{S. Dietrich}

\affiliation{Max-Planck-Institut f\"ur Intelligente Systeme, Heisenbergstr.~3, 
70569 Stuttgart, Germany}

\affiliation{IV. Institut f\"ur Theoretische Physik, Universit\"{a}t Stuttgart,
Pfaffenwaldring 57, D-70569 Stuttgart, Germany}

\date{\today}

\begin{abstract}
  We show theoretically that near a fluid-fluid interface a single
  active colloidal particle generating, e.g., chemicals or a
  temperature gradient experiences an effective force of hydrodynamic
  origin. This force is due to the fluid flow driven by Marangoni
  stresses induced by the activity of the particle; it decays very
  slowly with the distance from the interface, and can be attractive
  or repulsive depending on how the activity modifies the surface
  tension. We show that, for typical systems, this interaction can
  dominate the dynamics of the particle as compared to Brownian
  motion, dispersion forces, or self-phoretic effects. In the
    attractive case, the interaction promotes the self--assembly of
    particles into a crystal--like monolayer at the interface.
\end{abstract}



\maketitle

\label{Introduction}
Significant attention has been paid lately to micrometer sized
particles capable of self-induced motility
\cite{ebbens,SenRev,Sanchez2015}. They are seen as promising
candidates for novel techniques in chemical sensing \cite{RevLabChip}
or water treatment \cite{Soler2014}. The motion of active colloidal
particles has been the subject of numerous experimental
\cite{SenRev,ebbens,Golestanian2012,Fisher2014,Sanchez2015} and
theoretical
\cite{Golestanian2005,Julicher,Popescu2011EPL,Kapral2013,Lowen2011}
studies.  One realization is a particle with a catalytic surface
promoting a chemical reaction in the surrounding solution
\cite{Pine2013}.  For an axisymmetric particle lacking fore-and-aft
symmetry, the distributions of reactant and product molecules may
become non-uniform along its surface and the particle could move due
to self-induced phoresis \cite{Michelin2015}. If the particle is
spherically symmetric, it will remain immobile in bulk solution but
can be set into motion by the vicinity of walls or other particles
(not necessarily active) which break the spherical symmetry
\cite{Kapral2007,Popescu2011EPL,Pine2013,Michelin2015,Baraban2012}.

A relevant case corresponds to the movement of active particles
bounded by a fluid-fluid interface. This situation raises new issues,
in particular if the reactants or the products have a significant
effect on the properties of the fluid interface implying
\textit{tensioactivity}. For example, it has been recently predicted
that catalytically active, spherical particles which are trapped
\textit{at} the interface may be set into motion \textit{along} the
interface by Marangoni flows, self-induced via the spatially
non-uniform distribution of tensioactive molecules
\cite{Lauga2011,Stone2014,Masoud2014}.  (A similar motility mechanism
can originate from thermally induced Marangoni flows if, e.g., the
particle contains a metal cap which is heated by a laser beam
\cite{Wurger2014}.) Furthermore, self-induced Marangoni flows,
combined with a mechanism of triggering spontaneous
symmetry--breaking, have also been used to develop self-propelled
droplets \cite{Sugawara2006,Sugawara2007,Herminghaus2011}.

However, another category of experimental situations occurs if the
particles are not trapped \textit{at} the interface but may
\textit{reside in the vicinity} of the interface or \textit{get near}
it during their motion. In this study we provide theoretical evidence
that such catalytically active or locally heated spherical particles,
although immobile in bulk, experience a very strong, long-ranged
effective force field due to the Marangoni stresses self-induced at
the interface. This force of hydrodynamic origin manifests itself at
spatial length scales much larger than those of typical wetting
forces. It gives rise to a drift of the particle towards or away from
the fluid interface, depending only on how the tensioactive agent,
i.e., a gradient in chemical concentration or in temperature, affects
the interface. This effect dominates any possible self-phoresis or
dispersion interactions, and acts on time scales which can be orders
of magnitude shorter than those associated with Brownian diffusion.
This drift can facilitate particle adsorption towards the interface
and therefore has important implications for the self-assembly of
particles at fluid--fluid interfaces. We complement the theoretical
calculations with a thorough analysis regarding the observability of
these phenomena in future experiments.

\begin{figure}[t]
  \includegraphics[width=0.30\textwidth]{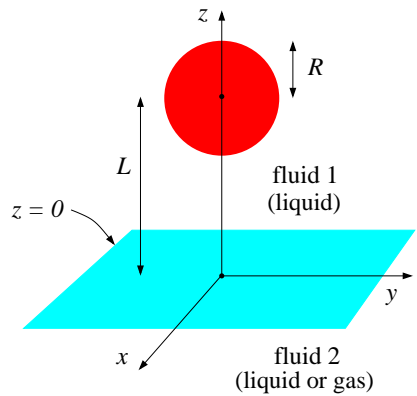}
  \caption{Coordinates and configuration of the system. The interface
    between fluid 1 (liquid) and fluid 2 (liquid or gas) is located at
    $z = 0$.  }
  \label{fig:setup}
\end{figure}

\label{Model}
The model system consists of a spherical colloidal particle with
radius $R$ in front of a flat interface at $z=0$ between two
immiscible fluids. Fluid 1 (2) occupies the half space $z>0$ ($z<0$)
(see \fig{fig:setup}). The spherical particle is located in fluid 1;
its center is at $\bx_0=(0,0,L)$ with $L > R$, i.e., the particle does
not penetrate through the interface. By virtue of a chemical reaction
occurring uniformly over its surface \footnote{The straightforward
  extension to Janus particles does not change the main conclusions as
  contained in \eq{eq:VlargeL}.}, the particle acts as a spherically
symmetric source (or sink) of a chemical species $A$. We assume that
the time scale for diffusion of $A$ is much shorter than any relevant
time scale associated with fluid flows \footnote{See the Supplementary
  Information for details}. Therefore we consider only the stationary
state neglecting advection by the ensuing Marangoni flow.
Additionally, the number density $c(\bx = \br + z \be_z)$ of species
$A$ (with $\br=(x,y,0)$ in the following) is assumed to be
sufficiently small so that for $A$ the ideal--gas approximation holds,
and thus $c(\bx)$ obeys Fick's law for diffusion with diffusivity
$D_\alpha$ in fluid $\alpha$ ($= 1\,, 2$):
\begin{subequations}
\label{eq:c}
\begin{equation}
  \label{eq:diff}
  \nabla^2 c(\bx) = 0 ,
  \qquad 
   \bx\in{\textrm{fluid 1 or 2}} ,
\end{equation}
subject to the boundary conditions \cite{Note2} that (i) a single
reservoir of species $A$ fixes the number density far away from the
particle to be $c_\alpha^\infty$ in fluid $\alpha$ (= 1, 2), (ii) the
discontinuity of $c(\bx)$ at the interface, given by $\lambda :=
c(\br, z = 0^-) / c(\br, z = 0^+) = c_2^\infty/c_1^\infty$, is
determined, as in equilibrium, by the distinct solvabilities of
species $A$ in the two fluids, (iii) the current of species $A$ along
the direction of the interface normal is continuous at the interface
($D_1 \,(\partial c/\partial z)|_{z=0^+} = D_2 \,(\partial c/\partial
z)|_{z=0^-}$), and (iv) the current at the surface $\mathcal{S}_p$ of
the particle is
\begin{equation}
  \label{eq:cflux}
   {\bn \cdot [-D_1 \nabla c(\bx)]} = \frac{Q}{4\pi R^2} ,
  \qquad
  \bx\in\mathcal{S}_p, 
\end{equation}
\end{subequations}
where $\bn$ is the unit vector normal to $\mathcal{S}_p$ (pointing into fluid 1); $Q > 0$ 
($Q < 0$) is the rate of production (annihilation) of species $A$.

The surface tension $\gamma$ of the fluid interface is assumed to
depend on the local number density of species $A$ and is modeled
within the local equilibrium approximation as \cite{Note2}
\begin{equation}
  \label{eq:gamma}
\gamma(\br) = \gamma_0 - b_0 \left[ c(\br, z=0^+)- c_1^\infty \right]\,.
\end{equation}
Here, $\gamma_0$ is the surface tension in equilibrium in which the density of $A$ 
in fluid 1 is $c_1^\infty$; the effect of local deviations thereof are quantified by
the coefficient $b_0$, the sign of which depends on the chemical.

This inhomogeneity of the surface tension induces Marangoni stresses which set 
the fluids into motion. The velocity field $\bv(\bx)$ can be derived as solution 
of the Stokes equations (i.e., in the limit of incompressible flow and 
negligible inertia):
\begin{subequations}
  \label{eq:v}
\begin{equation}
  \label{eq:stokes}
  \nabla\cdot\bv(\bx) = 0 ,
  \quad
  \nabla\cdot\overset{\leftrightarrow}{\sigma}(\bx) = 0 ,
  \qquad
  \bx \in {\textrm{fluid 1, 2}},
\end{equation}
where $\overset{\leftrightarrow}{\sigma}(\bx) = \eta(\bx) [ \nabla\bv +
  \left(\nabla\bv\right)^\dagger ] - p(\bx) \mathcal{I}$ is the stress tensor
in the fluid, $p(\bx)$ the pressure, $\eta(\bx)$ the viscosity, and $\mathcal{I}$ 
denotes the second--rank identity tensor. The Stokes equations are subject to 
the following boundary conditions: (i) vanishing velocity at infinity, (ii) no 
slip flow at the surface of the particle, (iii) at the interface, continuity of 
the tangential velocity and vanishing of the normal velocity, and (iv) balance 
between the tangential fluid stresses and the Marangoni stresses induced by the 
gradient of the surface tension along the interface:
\begin{equation}
  \label{eq:tangentstress}
  (\mathcal{I} - \be_z \be_z ) \cdot 
    \left[ 
      \left. \overset{\leftrightarrow}{\sigma}\right|_{z=0^+} - 
      \left.\overset{\leftrightarrow}{\sigma}\right|_{z=0^-}
    \right] \cdot\be_z
    = - \nabla_\parallel \gamma \,.
\end{equation}
\end{subequations}
The tensor $(\mathcal{I} - \be_z \be_z)$ provides the projection onto
the interfacial plane and $\nabla_\parallel =
(\partial_x,\partial_y,0)$ is the nabla operator within the
interfacial plane. (Actually, the interface must deform so that the
normal component of the fluid stresses can be balanced by the Laplace
pressure. \textit{A posteriori} it turns out \cite{Note2} that this
deformation is typically so small that the flat interface
approximation is reliable.)

\label{Results} 
The translation velocity $\boldsymbol{\mathcal{V}}$ of the particle,
or equivalently the force $\bF$ exerted by the particle on the fluid,
can be inferred from the Lorentz reciprocal theorem \cite{KiKa91}. We
consider the auxiliary flow field $\bv_\mathrm{aux}(\bx)$, for the
same geometrical setup, corresponding to the translation of a rigid,
spherical, \emph{chemically passive} (i.e., without Marangoni
stresses) particle in front of a flat fluid interface. This is the
solution of \eq{eq:stokes} subject to the same boundary conditions as
above but with $\nabla_\parallel \gamma = 0$ in \eq{eq:tangentstress},
a problem studied in Refs.~\cite{Bart68,LCL79,LeLe80}. We thus obtain
\cite{Note2}
\begin{eqnarray}
  \label{eq:faxen}
  \bF_\mathrm{aux}\cdot {\boldsymbol{\mathcal{V}}} - 
{\boldsymbol{\mathcal{V}}}_\mathrm{aux}\cdot\bF & = & 
  \int_{z=0} {d^2\br} \; \nabla_\parallel \gamma(\br) \cdot \bv_\mathrm{aux} (\br) 
  \nonumber \\
  & = & - \oint_{\mathcal{S}_p} {dS}\;
  \bn\cdot\overset{\leftrightarrow}{\sigma}_\mathrm{aux}(\bx) \cdot\bu(\bx)\,,
\end{eqnarray}
with 
\begin{subequations}
  \label{eq:flowField}
  \begin{equation}
    \label{eq:marangoni}
    \bu(\bx) = \int_{z=0} {d^2\br'}\; \nabla_\parallel 
\gamma({\br'})\cdot\mathcal{O}(\bx-{\br'}) 
  \end{equation}
in terms of the Oseen tensor,
  \begin{equation}
    \label{eq:oseen}
    \mathcal{O}(\bx) = \frac{1}{8\pi\eta_+ x}
    \left[ \mathcal{I} + \frac{\bx\bx}{x^2} \right] ,
    \quad
    \eta_+ := \frac{1}{2} (\eta_1+\eta_2) .
  \end{equation}
\end{subequations}
Here, $\bu(\bx)$ is the Marangoni flow, which would be induced solely
by the Marangoni stresses $\nabla_\parallel \gamma(\br)$, i.e., as if
the surface of the particle would not impose any boundary condition on
the flow \cite{Note2}.  Note that \eq{eq:faxen} can be interpreted as
a generalization of the Fax{\'e}n laws \cite{HaBr73} for the present
problem. For a force--free ($\bF=0$ in \eq{eq:faxen}) spherical
particle, the problem exhibits axial symmetry, which implies that
$\boldsymbol{\mathcal{V}}$ is parallel to $\be_z$. Thus, it suffices
to solve the auxiliary problem with $\bF_\mathrm{aux} \parallel
\be_z$, which allows one to introduce the dimensionless stream
function $\psi_\mathrm{aux}(r,z)$, in terms of which \eq{eq:faxen}
reduces to \cite{Note2}
\begin{equation}
  \label{eq:faxen2}
  {\boldsymbol{\mathcal{V}}} = - \be_z \, 2\pi R^2 b_0 \Gamma_z \int_0^\infty dr \; 
  \left.\frac{\partial c}{\partial r}\right|_{z=0^+}
  \left.\frac{\partial\psi_\mathrm{aux}}{\partial z}\right|_{z=0} ,
\end{equation}
where $\Gamma_z$ is the $L$--dependent mobility of a (chemically
passive) rigid spherical particle moving normal to the planar fluid
interface \cite{LeLe80}.

\begin{figure}[t]
   \includegraphics[width=0.45\textwidth]{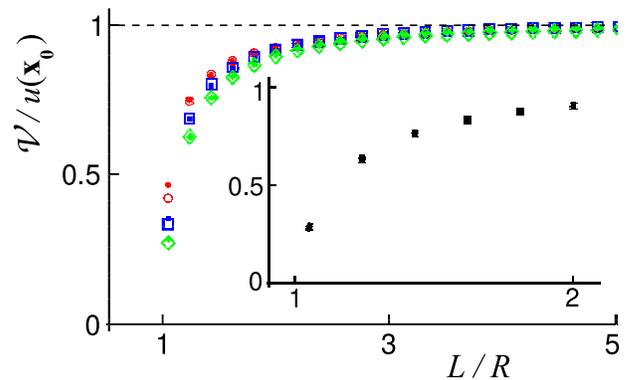}
   \caption{
     \label{fig:V}
     The ratio ${\mathcal{V}}/u(\bx_0)$ (\eqs{eq:faxen2} and
     (\ref{eq:VlargeL})) as a function of $L/R$ for $\lambda D_2/D_1 =
     0.1, 1, 10$ ($\Circle, \square, \Diamond$) and $\eta_2/\eta_1 =
     0.1, 10$ (open, filled). The inset provides an enlarged view of
     the range $L/R \lesssim 2$ for $\lambda D_2/D_1 = 10$ and
     $\eta_2/\eta_1 = 0.02, 0.1, 1, 10, 50$ ($\Circle, \square,
     \Diamond, \triangle, \times$); at that scale, the data are
     indistinguishable. The dashed line indicates the approximation
     provided by \eq{eq:VlargeL}.  }
\end{figure}

The boundary--value problems given by \eqs{eq:c} and (\ref{eq:v}),
subject to the coupling provided by \eq{eq:gamma}, can be solved
exactly as a series in terms of bipolar coordinates \cite{Note2}.
However, the relevant phenomenology can be highlighted and significant
physical intuition can be gained from an approximate closed form valid
asymptotically in the limit $R/L\to 0$. We therefore proceed with the
latter; its range of validity will be assessed later by comparison
with the exact solution (see, c.f., \fig{fig:V}).
\begin{figure}[t]
  \includegraphics[width=0.45\textwidth]{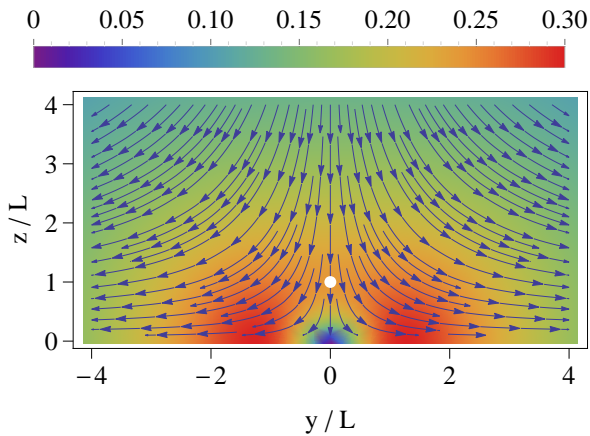}
  \caption{
Vertical cut through the Marangoni flow $\bu(\bx)$ in the limit $R/L\to 0$ 
(\eq{eq:uradsym2}). The streamlines follow the direction of the vector field 
(assuming $Q b_0 > 0$), while the color code corresponds to $|\bu(\bx)|$ in units of 
$|Q b_0|/(16\pi D_+ \eta_+ {L})$. The center of the particle (white dot) is 
at $y=0, z/L = 1$. The three-dimensional flow field is obtained by 
rotation around the $z$--axis and mirror reflection with respect to the interfacial plane 
$z=0$. (This flow is driven by the stress located \emph{at} the interface, not 
by the particle.)
\label{fig:flow} 
}
\end{figure}

To lowest order in $R/L$, the solution of \eq{eq:c} for the given boundary conditions can 
be obtained using the method of images in terms of monopoles located at $\bx_0=(0,0,L)$ 
and $\bx_0^*=(0,0,-L)$. In fluid 1 ($z > 0$) one has
\begin{equation}
  \label{eq:cmonopole}
  c(\bx) = c_1^\infty + \frac{Q}{4\pi D_1} \left[
    \frac{1}{|\bx-\bx_0|}
    + \frac{D_1- \lambda D_2}{D_1 + \lambda D_2} \frac{1}{|\bx-\bx_0^*|}
  \right]\,.
\end{equation}
Accordingly, the Marangoni flow (illustrated in \fig{fig:flow}) is given by 
Eqs.~(\ref{eq:gamma}) and (\ref{eq:marangoni}) as $\bu(\bx) = \be_z u_z (r,z) + 
\be_r u_r(r,z)$ with \cite{Note2}
\begin{subequations}
  \label{eq:uradsym2}
  \begin{equation}
    u_z (r, z) = - \frac{Q b_0}{16\pi D_+ \eta_+} 
    \frac{z (|z|+L)}{[ r^2 + (|z|+L)^2 ]^{3/2}} \,,
  \end{equation}
  \begin{equation}
    \label{eq:ur}
    u_r (r, z) = \frac{Q b_0}{16\pi D_+ \eta_+ r} 
    \left[ 1 -
      \frac{r^2 L + (|z|+L)^3}{[ r^2 + (|z|+L)^2 ]^{3/2}}
    \right]\,{,}
  \end{equation}
\end{subequations} 
and $D_+ := (D_1+\lambda D_2)/2$. The integral over $\mathcal{S}_p$ in the second 
line of \eq{eq:faxen} can be evaluated by expanding the Marangoni flow in terms of 
a Taylor series about the particle center so that asymptotically $\bu(\bx) \approx 
\bu(\bx_0)$ for $R/L \to 0$. Since $\oint_{\mathcal{S}_p} dS\; 
\bn\cdot\overset{\leftrightarrow}{\sigma}_\mathrm{aux}(\bx) = - \bF_\mathrm{aux}$, 
and $\bF_\mathrm{aux}$ can be chosen arbitrarily, one concludes that a force--free 
($\bF = 0$) active particle at a distance $L$ from the interface is carried by the 
flow with velocity
\begin{equation}
  \label{eq:VlargeL}
  \boldsymbol{\mathcal{V}}(L) 
  \approx \bu(\bx_0) \approx - \be_z \frac{Q b_0}{64\pi D_+ \eta_+ L} 
\end{equation}
due to the self-induced Marangoni stresses. Since $\boldsymbol{\mathcal{V}}~||~\be_z$, 
this implies a time dependence of $L$.

\label{Discussion}
Equation (\ref{eq:VlargeL}) captures the essence of all the relevant
phenomenology.  (i) For $b_0 > 0$ (i.e., the generic surfactant case)
the particle drifts towards or away from the interface if it is a
source ($Q > 0$) or a sink $(Q < 0)$ of species $A$, respectively. For
$b_0 < 0$, the behavior is reversed.  (ii) The slow $1/L$ decay of
$\mathcal{V}$ is tantamount to a long--ranged interaction with the
fluid interface. The associated phenomenology can dominate the
influence of dispersion forces between the particle and the interface,
which decay $\sim 1/L^4$ at best \cite{Parsegian_book}, and
also the motion due to Brownian diffusion alone.

The argument can be quantified by introducing the diffusion
coefficient $D_p : = k_B T/(6\pi\eta_+ R)$ of the particle in a medium
of viscosity $\eta_+$ at temperature $T$.  Equation~(\ref{eq:VlargeL})
leads to the Peclet number of the particle
\begin{equation}
  \label{eq:q}
  \mathrm{Pe}(L) := \frac{R \,|\boldsymbol{\mathcal{V}}(L)|}{D_p} = |q| \frac{R}{L} ,
  \qquad
  q := \frac{3 Q b_0 R}{32 D_+ k_B T} \, .    
\end{equation}
Dominance of drift means $\mathrm{Pe}(L) \gtrsim 1$. Thus, one
estimates the distance $L_\mathrm{max}$ from the interface, beyond
which the motion of the particle is controlled by diffusion rather
than by drift, as $\mathrm{Pe}(L_\mathrm{max}) = 1$ so that
$L_\mathrm{max} = |q| R$. Focusing on the case $b_0 >0$, one can
determine the time $t_\mathrm{drift}$ it takes the particle to reach
the interface starting (for $Q>0$) from a given distance $L_0$ (or,
for $Q<0$, to reach a given distance $L_0$ starting from near the
interface) via straightforward integration of the equation of motion
$dL/dt = \be_z \cdot {\boldsymbol{\mathcal{V}}}$ (within the
overdamped regime \cite{Note2}). This renders the drift time
$t_\mathrm{drift} = t_\mathrm{diff}/({2} |q|)$ in terms of the time of
diffusion $t_\mathrm{diff} := L_0^2/D_p$ over the same distance $L_0$.
Therefore, for large values of $|q|$ the drift caused by the Marangoni
flow, rather than diffusion, dominates the dynamics of the particle.

In order to estimate the magnitude of $q$, we shall use that typically
$b_0$ can take values in the range from $b_0 \sim -10^{-3} \;
{\mathrm{N/(m \,\times M)}}$ (M denotes mol/liter) for simple
inorganic salts in water \cite{JoRa41}, up to $b_0 \sim 10^2 \;
\mathrm{N/(m \,\times M)}$ for dilute solutions of surfactants (i.e.,
far from their critical micelle concentrations) \cite{HGC98}.  We
consider two distinct set-ups of potential experimental relevance.
\newline(i) The particle is a source. The chemical species $A$ is
molecular oxygen liberated from peroxide in aqueous solution by a
platinum--covered particle. For the experimental conditions described
in Ref.~\cite{PKOS04}, one has $Q/(4\pi R^2) \approx 10^{-3}\;
\mathrm{mol/(s\times m^2)}$ (compare \eq{eq:cflux}). At room
temperature ($300\;\mathrm{K}$) and for $R \simeq 1\, \mu\mathrm{m}$,
\eq{eq:q} leads to $|q| \sim 3 \times 10^{-4} \times
(D_+/(\mathrm{m}^{2} \times \mathrm{s}^{-1}))^{-1} (|b_0|/(\mathrm{N}
\times \mathrm{m}^{-1} \times \mathrm{M}^{-1}))$. For an air (fluid
2)--water (fluid 1) interface \footnote{It can be shown that
      the assumptions of incompressibility (low Mach number) and
      negligible inertia (low Reynolds number) hold also in the gas
      phase for the parameter values of interest.}, diffusion in the
gas phase dominates (typically $D_2 \simeq 10^{-5}\; \mathrm{m^2/s}$
and $D_1 \simeq 10^{-9}\; \mathrm{m^2/s}$ \cite{Ju1989}), and $D_+
\approx \lambda D_2/2 \simeq 10^{-4}\; \mathrm{m^2/s}$ (since $\lambda
\sim 10-100$ for oxygen and air--water interface \cite{Battino1983}),
while $b_0$ is in the lower range of values \cite{PKOS04}. Thus
$L_\textrm{max}/R = |q| \sim 10^{-2}$, which explains the lack of
reports of such effects for experimental set-ups as in
Ref.~\cite{PKOS04}. However, for a liquid--liquid interface (e.g.,
water--decane), one has $D_+ \simeq D_1 \simeq D_2 \simeq 10^{-9}\;
\mathrm{m^2/s}$ (one expects $\lambda \lesssim 1$ \cite{Ju1989}) and
thus $|q| \gtrsim 10^{2}$ across the range of values $b_0$ noted
above. Therefore, for the same experimental set-ups of active
colloids, but which involve liquid-liquid interfaces instead of
liquid-gas ones, we predict that the effective interactions discussed
here dominate. The same conclusion holds for liquid--gas interfaces
but with a reaction product with very low solubility in the gas phase
(i.e., $\lambda \ll 1$).  \newline (ii) The particle is a sink, i.e.,
the tensioactive species $A$ is absorbed completely by the particle.
One can infer $Q$ from the diffusion--limited regime in which the
surface of the particle acts as an absorbing boundary so that
$c(\bx\in\mathcal{S}_p) = 0$, which with \eq{eq:cmonopole} provides
the estimate $Q \approx -4\pi D_1 R c_1^\infty$ as $R/L\to 0$. With
\eq{eq:q} one arrives at $|q| \sim 3 \times 10^8 \times (D_1/D_+)
(R/\mu\mathrm{m})^2 (c_1^\infty/\mathrm{M}) (|b_0|/(\mathrm{N} \times
\mathrm{m}^{-1} \times \mathrm{M}^{-1}))$.  Therefore,
$L_\textrm{max}/R = |q|$ can indeed be large for colloidal particles
even if species $A$ is only weakly tensioactive (i.e., $|b_0|$ small)
and even for liquid--gas interfaces ($D_1/D_+ \ll 1$).

The availability of an exact series representation for $\mathcal{V}$
as given by \eq{eq:faxen2} allows us to assess the range of validity
of the asymptotic approximation (\eq{eq:VlargeL}) discussed above. For
several values of the viscosity and diffusivity ratios \cite{Note2},
\fig{fig:V} shows ${\mathcal{V}}/u(\bx_0)$ as a function of the
separation $L/R$. It turns out that $u(\bx_0)$ provides a reliable
approximation (less than 10\% relative error) of the exact solution
down to separations $L/R \simeq 2$, i.e., covering most of the range
within which the model is relevant. Furthermore, the deviations from
$u(\bx_0)$ depend very weakly on the ratios $\eta_2/\eta_1$ and
$\lambda D_2/D_1$.

\label{Conclusions}
These results, yet to be explored experimentally, have several
implications, which we highlight in conclusion. First, as noted in
the Introduction, if the particle maintains a temperature gradient,
e.g., through local heating, the very same equations hold with the
temperature playing the role of the number density $c(\bx)$.
Therefore, all the phenomenology discussed above extends to this case,
too.
Second, sufficiently close to the interface the drift due to the
induced Marangoni flows can dominate even self-phoretic motion. For
instance, a Janus particle of size $R = 1~\mathrm{\mu m}$ (with
$D_p\sim 10^{-13} \;\mathrm{m}^2/\mathrm{s}$ in water) and
self-propelling with a typical velocity of $\sim 1~\mathrm{\mu m/s}$
\cite{ebbens} has a Peclet number $\mathrm{Pe}_{\mathrm{phor}} \simeq
10$, which, e.g., at distances $L/R < 10$ is smaller than
$\mathrm{Pe}(L) = (R/L)|q|$ if $|q|~\gtrsim~10^2$ (see \eq{eq:q}).
Third, based on the single-particle phenomenology studied here one can
infer potentially significant collective effects. Consider for example
a dilute suspension of active particles which are driven towards the
interface by the Marangoni stresses and in addition experience a
short-ranged repulsion by the interface (e.g., due to electrostatic
double layer interactions). 
Then the particles are expected to reside near the interface while
experiencing a mutual long-ranged lateral repulsion as each particle
is carried by the Marangoni flows induced by the others (see the flow
lines in \fig{fig:flow} and \eq{eq:ur}, which tells that $u_r$
exhibits also a slow in-plane decay $\sim 1/r$). Therefore, near the
interface and in the presence of lateral boundaries self-organized
crystal-like monolayers could be reversibly assembled and
``dissolved'' by simply turning on and off the activity of the
particles. 
Finally, we note that this effective lateral pair interaction violates
the action-reaction principle because non--identical particles (e.g.,
due to size-polydispersity, different production rates $Q$, or a
heterogeneous coverage of the surface) create Marangoni flows of
different strength. As for other systems in which such violations
occur \cite{Lowen2015}, this feature can be expected to give rise to a
complex collective behavior and to a rich, barely explored
phenomenology. 

\begin{acknowledgments}
\label{Acknowledgments}
A.D. acknowledges support by the Spanish Government through Grant
FIS2011-24460 (partially financed by FEDER funds).
\end{acknowledgments}


%

\end{document}


\title{SUPPLEMENTARY INFORMATION}

\maketitle

\tableofcontents

\section{The boundary--value problem posed in \eq{eq:c}}

With the hypothesis that the time scale for diffusion of species $A$ is sufficiently 
short (see Sec.~\ref{sec:approx}), the stationary spatial distribution $c(\bx)$ of 
species $A$ can be obtained by assuming local equilibrium. This allows one to define a 
local free energy, depending on temperature and the local number density $c(\bx)$, 
as well as the corresponding chemical potential $\mu(c(\bx))$. (Here, the 
temperature dependence of $\mu$ is not explicitly indicated because we shall focus 
on isothermal systems.) The chemical potential can be decomposed, as usual, into an ideal 
gas contribution and an excess part which includes, among other terms, the potential 
energy $U_\alpha$ per $A$ molecule immersed in the background fluid $\alpha$ 
(i.e., the solvability of $A$ in $\alpha$). Since the latter depends on the fluid 
$\alpha$, we write the chemical potential $\mu(c(\bx))$ as $\mu_\alpha(c(\bx))$, 
$\alpha=1,2$. Therefore, $c(\bx)$ is the  solution of the following boundary 
value problem:
  \begin{subequations}
    \label{eq:cproblem}
    \begin{equation}
      \label{eq:fick}
      \nabla\cdot\mathbf{j}_\alpha = 0 , 
      \quad \mathbf{j}_\alpha (\bx) := - \Gamma_\alpha c(\bx) \nabla \mu_\alpha (c(\bx)) ,
      \qquad
      \mathrm{if}\;\bx\in\mathrm{fluid}\;\alpha=1,2\,,
    \end{equation}
    \begin{equation}
      \label{eq:mu0}
      \mu_\alpha(c(\bx))\to\mu_0 , 
      \qquad
      \mathrm{if}\;|\bx|\to\infty ,
    \end{equation}
    \begin{equation}
      \label{eq:cjump}
      \mu_1(c(\br,z=0^+)) = \mu_2(c(\br,z=0^-)) ,
      \qquad
      \textrm{at the interface $z=0$},
    \end{equation}
    \begin{equation}
      \label{eq:normalj}
      \be_z\cdot\left[ \mathbf{j}_1(\br,z=0^+) - \mathbf{j}_2(\br,z=0^-) \right] = 0\,,
      \qquad
      \textrm{at the interface $z=0$},
    \end{equation}
    \begin{equation}
      \label{eq:Qflux}
      \bn \cdot \mathbf{j}_1(\bx) = \frac{Q}{4\pi R^2} \,,
      \qquad
      \textrm{if } \bx\in\mathcal{S}_p.
    \end{equation}
  \end{subequations}
Here, \eq{eq:fick} is Fick's law for the particle current density $\mathbf{j}(\bx)$, 
expressed in terms of the chemical potential and the mobility $\Gamma_\alpha$ of 
species $A$ in the fluid $\alpha$. \eq{eq:mu0} fixes the chemical potential of species
$A$ far from the colloidal particle to be that of the reservoir, i.e., $\mu_0$. This 
determines the number densities $c_1^\infty$ and $c_2^\infty$ in the fluids far from 
the particle: 
  \begin{equation}
    \label{eq:cinfty}
    \mu_1(c_1^\infty) = \mu_2(c_2^\infty) = \mu_0 .
  \end{equation}
Equation~(\ref{eq:cjump}) represents the condition of (local) equilibrium at the 
interface and determines the jump in $c(\bx)$ at the interface. 
Equation~(\ref{eq:normalj}) expresses the continuity of the normal component of the 
current of $A$ molecules through the interface, so that there is no accumulation of 
species $A$ there. Finally, \eq{eq:Qflux} accounts for the colloidal particle being a 
source (or sink) of species $A$. 

In order to be able to obtain an analytical solution of this boundary--value problem 
a specific expression for the chemical potential of species $A$ as a function of 
the number density $c$ is needed. For reasons of simplicity, we consider 
the dilute limit, so that for $A$ the ideal--gas approximation can be 
used. This implies
  \begin{equation}
    \label{eq:muideal}
    \mu_\alpha(c) = \kt \ln \frac{c}{c_0} + U_\alpha ,
    \qquad
    \alpha=1,2.
  \end{equation}
Here, $c_0$ is a constant reference number density and $U_\alpha$ is the effective 
potential energy of a molecule $A$ immersed in the fluid $\alpha$. A nonzero value 
of the difference $U_1-U_2$ indicates different solvabilities of $A$ in the 
two fluids. With this expression for the chemical potential, the current 
density takes the form
    \begin{equation}
      \mathbf{j}_\alpha(\bx) = - D_\alpha \nabla c(\bx) ,
      \qquad
      D_\alpha := \kt \Gamma_\alpha, 
    \end{equation}
where $D_\alpha$ is the diffusivity of species $A$ in fluid $\alpha$, which 
in the dilute limit is spatially constant. Correspondingly, the boundary--value 
problem~(\ref{eq:cproblem}) reduces to
  \begin{subequations}
    \label{eq:cproblemideal}
    \begin{equation}
      \label{eq:cnabla2}
      \nabla^2 c = 0 , 
      \qquad
      \mathrm{if}\;\bx\in\textrm{fluid 1 or 2},
    \end{equation}
    \begin{equation}
      \label{eq:mu02}
      \lim_{|\bx|\to\infty} c(\bx) = \left\{
        \begin{array}[c]{cl}
          c_1^\infty := c_0 \mathrm{e}^{(\mu_0-U_1)/\kt}, & z>0, \\
          & \\
          c_2^\infty := c_0 \mathrm{e}^{(\mu_0-U_2)/\kt} , & z<0, \\
        \end{array}
      \right.
    \end{equation}
    \begin{equation}
      \label{eq:cjump2}
      \lambda := \frac{c(\br,z=0^-)}{c(\br,z=0^+)} = \mathrm{e}^{(U_1-U_2)/\kt} 
      \;\; 
      = \frac{c_2^\infty}{c_1^\infty}\,, 
      \qquad
      \textrm{at the interface $z=0$},
    \end{equation}
    \begin{equation}
      \label{eq:normalj2}
      D_1 \left.\frac{\partial c}{\partial z}\right|_{z=0^+} =
      D_2 \left.\frac{\partial c}{\partial z}\right|_{z=0^-} ,
      \qquad
      \textrm{at the interface $z=0$},
    \end{equation}
    \begin{equation}
      \label{eq:Qflux2}
      \bn \cdot [-D_1 \nabla c(\bx)] = \frac{Q}{4\pi R^2} ,
      \qquad
      \textrm{if } \bx\in\mathcal{S}_p,
    \end{equation}
  \end{subequations}
where we have introduced the ratio $\lambda$ which characterizes the 
discontinuity of the field $c(\bx)$ at the interface (above the molecular scale). In 
this manner, \eq{eq:c} and the corresponding boundary conditions (i--iv) in the main 
text are obtained. The exact solution of this system of equations will be provided in  
Sec.~\ref{sec:cbipolar}.

\section{The dependence of the surface tension on the field $c(\bx)$}

The surface tension $\gamma$ of the fluid interface depends on the structural 
properties of both fluid phases, in particular on the number density of species $A$ in 
each of them. In the absence of particle activity (i.e., $Q=0$ in \eq{eq:Qflux}), the 
solution of \eq{eq:cproblem} is the equilibrium state with
  \begin{equation}
    \label{eq:muequil}
    \mu_1(c(\bx))=\mu_2(c(\bx))=\mu_0, 
    \quad 
    \forall\bx
    \qquad\Rightarrow\qquad
    c(\bx) = \left\{
      \begin{array}[c]{cl}
        c_1^\infty, & \bx\in\mathrm{fluid}\;1, \\
        & \\
        c_2^\infty, & \bx\in\mathrm{fluid}\;2, \\
      \end{array}
    \right.
  \end{equation}
which is a homogeneous distribution of species $A$ in each fluid. The surface 
tension of the interface in this state is denoted as $\gamma_0$. A variation 
$\delta\mu_0$ of the chemical potential of the reservoir induces changes $\delta c_1$ 
and $\delta c_2$ of the number densities of $A$ in the two fluids. They follow from 
\eqs{eq:muequil} and (\ref{eq:cinfty}) as
  \begin{equation}
    \label{eq:deltamu}
    \delta c_1 \frac{\partial\mu_1}{\partial c_1}\left(c_1=c_1^\infty\right)
    =
    \delta c_2 \frac{\partial\mu_2}{\partial c_2}\left(c_2=c_2^\infty\right) ,
  \end{equation}
by retaining only the lowest order terms in the deviations. The corresponding change in 
the surface tension is taken into account consistently to lowest order in the 
deviations:
  \begin{equation}
    \label{eq:gammaeq}
    \gamma = \gamma_0 - b_0 \delta c_1 =  \gamma_0 - \hat{b}_0 \delta c_2 ,
    \qquad
    b_0 := -\frac{d\gamma}{d c_1} \left(c_1=c_1^\infty\right) ,
    \quad
    \hat{b}_0 := -\frac{d\gamma}{d c_2} \left(c_2=c_2^\infty\right) .
  \end{equation}
Here, the coefficient $b_0$ quantifies the effect of the deviations in $c_1$ and it 
encodes the effective molecular interactions between the $A$ molecules and the 
interface between the two fluids. We note that the variations in $\gamma$ 
can be expressed equivalently in terms of quantities pertaining only to fluid 2, 
because the equilibrium condition determines unambiguously the relationship between 
the densities of $A$ on each side of the interface: from \eq{eq:deltamu} one has the 
relationship
  \begin{equation}
    \frac{1}{b_0} \frac{\partial\mu_1}{\partial c_1}\left(c_1=c_1^\infty\right) 
    = \frac{1}{\hat{b}_0} \frac{\partial\mu_2}{\partial c_2}\left(c_2=c_2^\infty\right) .
  \end{equation}

If the particle is active, the distribution of species $A$ is described by a 
nonequilibrium, spatially varying state. In such a case, the surface tension of 
the interface is determined by considering local equilibrium: \eq{eq:gammaeq} holds 
but with the deviation $\delta c_1$ given locally by the difference 
$c(\br,z=0^+)-c_1^\infty$. This yields \eq{eq:gamma} in the main text.


\section{Derivation of Eqs.~(\ref{eq:faxen}-\ref{eq:faxen2}) in the main text}

The Lorentz reciprocal theorem \cite{KiKa91,Teubner1982} for the Stokes 
equation in the domain $\mathcal{D}$ outside the particle, which is bounded by the 
surface $\mathcal{S}_p$ of the particle and a surface $\mathcal{S}_\infty$ at infinity, 
takes the form
\begin{equation}
  \label{eq:lorenz}
  \oint_{{\mathcal{S}_p}\cup\mathcal{S}_\infty} dS\; {\bn_\mathrm{out}}\cdot 
  \overset{\leftrightarrow}{\sigma}_\mathrm{aux} \cdot \bv
  - \int_\mathcal{D} d^3\bx\; (\nabla\cdot\overset{\leftrightarrow}{\sigma}_\mathrm{aux})
  \cdot \bv   
  = \oint_{{\mathcal{S}_p}\cup\mathcal{S}_\infty} dS\; {\bn_\mathrm{out}} \cdot 
  \overset{\leftrightarrow}{\sigma} \cdot \bv_\mathrm{aux}
  - \int_\mathcal{D} d^3\bx\; (\nabla\cdot\overset{\leftrightarrow}{\sigma}) 
  \cdot \bv_\mathrm{aux} ,
\end{equation}
where $\bn_\mathrm{out}$ denotes the unit vector normal to the
bounding surface and oriented towards the exterior of the fluid (the
so-called \textit{outer normal}). The flow field $\bv(\bx)$ in
  \eq{eq:lorenz} corresponds to our model describing an active rigid
particle which moves with velocity $\boldsymbol{\mathcal{V}}$ in front of a 
fluid interface (in the Monge representation $z=f(x,y)$ characterized by a function 
$f(\br)$ of the lateral coordinates $\br:=(x,y,0)$) with a local surface tension 
$\gamma(\br)$, and is thus given as the solution of the boundary-value problem 
\begin{subequations}
  \begin{equation}
    \label{eq:stokesSInonflat}
    \nabla\cdot\bv = 0 ,
    \qquad
    \nabla\cdot\overset{\leftrightarrow}{\sigma} 
    = - \delta(z -f(x,y)) 
    \left[ \nabla_\parallel \gamma + 2 \gamma H \be_n \right] ,
    \qquad
    \mathrm{if}\;\bx\in\mathcal{D} ,
  \end{equation}
  \begin{equation}
    \bv(\bx) \textrm{ continuous everywhere,}
  \end{equation}
  \begin{equation}
    \label{eq:impermnonflat}
    \be_n \cdot\bv = 0 \textrm{ at the interface $z=f(x,y)$,}
  \end{equation}
  \begin{equation}
    \bv(\bx) = {\boldsymbol{\mathcal{V}}},
    \qquad
    \mathrm{if}\;\bx\in\mathcal{S}_p ,
  \end{equation}
  \begin{equation}
       \bv(|\bx|\to\infty) = 0\, ,
  \end{equation}
  \begin{equation}
    \overset{\leftrightarrow}{\sigma} := \eta(\bx) \left[ 
      \nabla\bv + \left(\nabla\bv\right)^\dagger 
    \right] - p \mathcal{I} .
  \end{equation}
\end{subequations}
With $p=f_x$, $q=f_y$, $r=f_{xx}$, $s=f_{xy}$, and $t=f_{yy}$, here $\be_n(x,y) := 
(-p,-q,1)/\sqrt{1+p^2+q^2}$ denotes the local unit vector normal to the interface and 
$H(x,y)=[r (1+q^2)-2 p q s + t (1+p^2)]/[2 (1+p^2+q^2)^{3/2}]$ is its local mean 
curvature. The effect of the interface on the flow is incorporated through 
\eq{eq:stokesSInonflat} (via the Marangoni stress $\nabla_\parallel \gamma$ and the 
Laplace pressure $2 \gamma H$, respectively) and \eq{eq:impermnonflat} (via the 
kinematic constraint of a stationary interface). The hydrodynamic flow deforms the shape 
of the interface relative to its flat configuration such that the Laplace pressure 
$2\gamma H$ due to the deformation can balance the pull by the flow, i.e., the difference 
in normal stresses across the interface. Thus the deformation $f(x,y)$ would be 
computed as part of the solution of the boundary--value problem. However, the relatively 
large value of the surface tension in typical set-ups implies that the deformation 
is usually small and the interface is close to the plane $z = 0$ everywhere. Therefore 
one can approximate the initial problem by assuming the interface to be flat
in the geometric sense (formally $f\to 0$, which means that the deformations are 
negligible compared to all other relevant length scales), but subject to an 
in--plane stress given by $\nabla_\parallel \gamma (\br)$ and subject to an 
unknown areal density of force $\be_z \phi(\br)$ which imposes the flat--interface 
constraint. The latter can be interpreted as the formal limit $2\gamma(\br) H(\br) 
\be_n \to \phi(\br) \be_z$, $\phi(\br)\; \mathrm{finite}$, of the Laplace pressure, 
if $H \to 0$ and $\gamma\to\infty$. This leads to the simplified boundary-value 
problem
\begin{subequations}
  \label{eq:fullflow}
  \begin{equation}
    \label{eq:stokesSI}
    \nabla\cdot\bv = 0 ,
    \qquad
    \nabla\cdot\overset{\leftrightarrow}{\sigma} 
    = - \delta(z) 
    \left[ \nabla_\parallel \gamma + \be_z \phi \right] ,
    \qquad
    \mathrm{if}\;\bx\in\mathcal{D} ,
  \end{equation}
  \begin{equation}
    \bv(\bx) \textrm{ continuous everywhere,}
  \end{equation}
  \begin{equation}
    \label{eq:imperm}
    \be_z\cdot\bv = 0 \textrm{ at the interface $z=0$},
  \end{equation}
  \begin{equation}
    \label{eq:V}
    \bv(\bx) = {\boldsymbol{\mathcal{V}}},
    \qquad
    \mathrm{if}\;\bx\in\mathcal{S}_p ,
  \end{equation}
  \begin{equation}
     \bv(|\bx|\to\infty) = 0\, ,
  \end{equation}
  \begin{equation}
    \overset{\leftrightarrow}{\sigma} := \eta(\bx) \left[ 
      \nabla\bv + \left(\nabla\bv\right)^\dagger 
    \right] - p \mathcal{I} ,
  \end{equation}
\end{subequations}
which is the boundary-value problem formulated in \eq{eq:v} in the main text. 
(In particular, \eq{eq:tangentstress} is obtained by integrating \eq{eq:stokesSI} over an 
infinitesimal cylindrical box (centered at the interface with its axis parallel to the 
interface normal $\be_z$), applying Gauss' theorem, and then taking the in--plane 
projection of the result.) This problem can be solved without the need to apply the stress 
balance boundary condition at the interface in normal direction arising from 
\eq{eq:stokesSI}, because the interface shape is prescribed and fixed to be planar. 
The constraining force $\phi$ follows from the discontinuity in the normal stress at the 
interface \emph{after} the solution has been found. Then $\phi$ can be used to estimate 
the curvature as $H(\br)=\phi(\br)/(2\gamma(\br))$. This allows one to check \emph{a 
posteriori} the self--consistency of the flat--interface approximation (see 
Sec.~\ref{sec:approx}).

As for $\bv_\mathrm{aux}(\bx)$, we choose it to be the flow field for the same geometry 
(within the same flat--interface approximation discussed above) but for a passive 
particle, i.e., in the absence of Marangoni flow. Thus $\bv_\mathrm{aux}(\bx)$ is the 
solution of \eq{eq:fullflow} but with $\nabla_\parallel \gamma=0$:
\begin{subequations}
  \label{eq:aux}
  \begin{equation}
  \label{eq:stokesSI_vaux}
    \nabla\cdot\bv_\mathrm{aux} = 0 ,
    \qquad
    \nabla\cdot\overset{\leftrightarrow}{\sigma}_\mathrm{aux} 
    = - \delta(z) \be_z \phi_\mathrm{aux},
    \qquad
    \mathrm{if}\;\bx\in\mathcal{D} ,
  \end{equation}
  \begin{equation}
    \bv_\mathrm{aux}(\bx) \textrm{ continuous everywhere,}
  \end{equation}
  \begin{equation}
    \label{eq:impermaux}
    \be_z\cdot\bv_\mathrm{aux} = 0 \textrm{ at the interface $z=0$},
  \end{equation}
  \begin{equation}
    \label{eq:Vaux}
    \bv_\mathrm{aux}(\bx) = {{\boldsymbol{\mathcal V}}}_\mathrm{aux} ,
    \qquad
    \mathrm{if}\;\bx\in\mathcal{S}_p ,
  \end{equation}
  \begin{equation}
    \bv_\mathrm{aux}(|\bx|\to\infty) = 0\,,
  \end{equation}
  \begin{equation}
    \overset{\leftrightarrow}{\sigma}_\mathrm{aux} := \eta \left[ 
      \nabla\bv_\mathrm{aux} + \left(\nabla\bv_\mathrm{aux}\right)^\dagger 
    \right] - p_\mathrm{aux} \mathcal{I} .
  \end{equation}
\end{subequations}
The exact solution of these equations for $\boldsymbol{\mathcal{V}} = \be_z \mathcal{V}$ 
will be provided in Sec.~\ref{sec:psiaux}. Since both fields $\bv(\bx)$ and 
$\bv_\mathrm{aux}(\bx)$ decay asymptotically at least as $1/x$ for $|\bx|\to\infty$, in 
\eq{eq:lorenz} the contribution from the surface $\mathcal{S}_\infty$ vanishes. The 
remaining terms can be evaluated easily. With \eq{eq:Vaux} one has
\begin{equation}
  \oint_{\mathcal{S}_p} dS \; 
  \bn_\mathrm{out}\cdot \overset{\leftrightarrow}{\sigma} \cdot \bv_\mathrm{aux} 
  = \left( - \oint_{\mathcal{S}_p} dS \; 
    \bn\cdot \overset{\leftrightarrow}{\sigma} \right) \cdot 
  {\boldsymbol{\mathcal V}}_\mathrm{aux} 
  = \bF \cdot {\boldsymbol{\mathcal V}}_\mathrm{aux}
\end{equation}
in terms of the force $\bF$, which the particle exerts on the fluid, and with the 
unit normal $\bn=-\bn_\mathrm{out}$ \textit{oriented into the fluid} (as defined in the 
main text). Similarly, due to \eq{eq:V} one has
\begin{equation}
  \label{eq:faux}
  \oint_{{\mathcal{S}_p}} dS \; 
  {\bn_\mathrm{out}}\cdot \overset{\leftrightarrow}{\sigma}_\mathrm{aux} \cdot \bv
  = \bF_\mathrm{aux} \cdot {\boldsymbol{\mathcal{V}}} \,.
\end{equation}
Since at the interface the flow $\bv_\mathrm{aux}$ has no vertical component (see 
\eq{eq:impermaux}), the application of \eq{eq:stokesSI} gives
\begin{equation}
  \int_\mathcal{D} d^3\bx\; (\nabla\cdot\overset{\leftrightarrow}{\sigma})
  \cdot \bv_\mathrm{aux} 
  = - \int_{z=0} {d^2\br}
  \; \nabla_\parallel\gamma (\br) \cdot \bv_\mathrm{aux}(\br) .
\end{equation}
Similarly, \eq{eq:stokesSI_vaux} leads to
\begin{equation}
  \int_\mathcal{D} d^3\bx\; (\nabla\cdot\overset{\leftrightarrow}{\sigma}_\mathrm{aux})
  \cdot \bv = 0 .
\end{equation}
Inserting these expressions into \eq{eq:lorenz}, one finds
\begin{equation}
  \label{eq:faxenSI}
  \bF_\mathrm{aux}\cdot {\boldsymbol{\mathcal{V}}} = \bF \cdot  
{\boldsymbol{\mathcal{V}}}_\mathrm{aux}
  + \int_{z=0} {d^2\br}\; \nabla_\parallel\gamma (\br) \cdot \bv_\mathrm{aux}(\br) ,
\end{equation}
which is the first line of \eq{eq:faxen} in the main text.

Turning now to the second line of \eq{eq:faxen}, we note that the solution 
$\bv_\mathrm{aux}({\bx'})$ can be represented as (see Eq.~(34) in Ref.~\cite{SePa11_1})
\begin{equation}
  \label{eq:green}
  \bv_\mathrm{aux}({\bx'}) = - \oint_{\mathcal{S}_p} dS\; 
  {\bn}\cdot\overset{\leftrightarrow}{\sigma}_\mathrm{aux}({\bx})\cdot 
  \mathcal{O}_\mathrm{aux}({\bx',\bx}) ,
\end{equation}
where the non--symmetric, 2nd--rank tensor $\mathcal{O}_\mathrm{aux}$ is the Green's 
function of the auxiliary problem, which can be constructed as a superposition of certain 
singularity solutions and their mirror images with respect to the plane $z=0$ 
\cite{JFD75,LCL79}. In particular, if the point $\bx'$ is located at the fluid interface, 
i.e., $\bx' = \br' = (x',y',0)$, this tensor takes the simple form (see Eq.~(25) 
in Ref.~\cite{SePa11_1})
\begin{equation}
  \mathcal{O}_\mathrm{aux}({\br',\bx}) 
  = \mathcal{O} (\br'-\bx) \cdot (\mathcal{I} - \be_z\be_z) 
\end{equation}
in terms of the (symmetric) Oseen tensor defined in \eq{eq:oseen} in the main text, 
representing the Green's function for the Stokes equation in an unbounded, homogeneous 
medium. Therefore, after inserting the integral representation of \eq{eq:green} into 
\eq{eq:faxenSI} and interchanging the integrations, one obtains
\begin{eqnarray}
  \int_{z=0} {d^2\br'}\; \nabla_\parallel\gamma ({\br'}) \cdot \bv_\mathrm{aux}
({\br'}) 
  & = & { - \oint_{\mathcal{S}_p} dS\;
    \bn\cdot\overset{\leftrightarrow}{\sigma}_\mathrm{aux}(\bx) \cdot 
    \int_{z=0} d^2\br' \; \mathcal{O} (\br'-\bx) \cdot (\mathcal{I} - \be_z\be_z)
    \cdot \nabla_\parallel\gamma (\br')} 
  \nonumber \\
  & = & - \oint_{\mathcal{S}_p} dS\; 
  {\bn}\cdot\overset{\leftrightarrow}{\sigma}_\mathrm{aux}({\bx}) \cdot \bu({\bx})
\end{eqnarray}
due to $\be_z\cdot\nabla_\parallel\gamma=0$, with the Marangoni flow $\bu(\bx)$ 
defined in \eq{eq:marangoni} in the main text.

In order to obtain \eq{eq:faxen2} in the main text, we notice that the symmetry of 
the problem warrants that the translational velocity ${\boldsymbol{\mathcal{V}}}$ of 
a force-free ($\bF=0$) moving active particle has only a component in 
the vertical direction. Therefore, for the application of \eq{eq:faxenSI} it suffices to 
solve the auxiliary problem with a force $\bF_\mathrm{aux}$ acting in the vertical 
direction. In such a case, $\bv_\mathrm{aux}(\bx=\br+z\be_z)$ is an axisymmetric
flow, which can be represented by means of a stream function  $\Psi_\mathrm{aux}(\bx) = 
\mathcal{V}_\mathrm{aux} R^2 \psi_\mathrm{aux}(\bx)$
as \cite{HaBr73}
\begin{equation}
  \label{eq:stream}
  \bv_\mathrm{aux}(\bx) = \frac{\mathcal{V}_\mathrm{aux} R^2}{r} \left[ 
    \frac{\br}{r} \frac{\partial\psi_\mathrm{aux}}{\partial z}
    - \be_z \frac{\partial\psi_\mathrm{aux}}{\partial r}
  \right] ,
\end{equation}
where $\psi_\mathrm{aux}{(r = |\br|,z)}$ is dimensionless, $\mathcal{V}_\mathrm{aux} = 
\Gamma_z \be_z\cdot\bF_\mathrm{aux}$, and $\Gamma_z$ is the $L$--dependent vertical 
mobility of a (chemically passive) rigid spherical particle in front of a planar 
fluid-fluid interface \cite{LeLe80} (see the next Sec. concerning the explicit 
expressions for $\psi_\mathrm{aux}$ and $\Gamma_z$). Thus \eq{eq:faxen} in the main text 
reduces to \eq{eq:faxen2} after the gradient of the surface tension is replaced by that of 
the concentration via \eq{eq:gamma}.

\section{Derivation of \eqs{eq:cmonopole} and (\ref{eq:uradsym2}) in the main 
text}

Equations~(\ref{eq:cmonopole}) and (\ref{eq:uradsym2}) represent the solution of the 
differential equations for the Marangoni flow in the asymptotic limit $R/L\to 
0$. First, we consider the boundary-value problem in \eq{eq:cproblemideal}: in this 
limit, one can approximate the colloidal particle by a point source. Thus
Eqs.~(\ref{eq:cnabla2}) and (\ref{eq:Qflux2}) can be combined into the single 
equation
  \begin{equation}
    \nabla^2 c = -\frac{Q}{D_1} \delta(\bx-\bx_0) ,
  \end{equation}
where $\bx_0=(0,0,L)$ is the position of the center of the particle. The solution of this 
differential equation, which satisfies the boundary condition in \eq{eq:mu02} at 
infinity, can be constructed via the method of images by making the ansatz
  \begin{subequations}
    \begin{equation}
      c(\bx) = c_1^\infty + \frac{1}{4\pi D_1} \left[ \frac{Q}{|\bx-\bx_0|}
        + \frac{Q'}{|\bx-\bx_0^*|} \right] ,
      \qquad
      z>0 ,
    \end{equation}
    \begin{equation}
    \label{eq:Q''}
      c(\bx) = c_2^\infty + \frac{Q''}{4\pi D_1 |\bx-\bx_0|} ,
      \qquad
      z<0 ,
    \end{equation}
  \end{subequations}
where $\bx_0^*=(0,0,-L)$ is the mirror point of the center of the particle with respect to 
the interface; $Q'$ and $Q''$ represent image sources. (Note that $Q''$ is defined such 
that \eq{eq:Q''} contains as a prefactor $1/D_1$ even for $z < 0$.) Their values are 
determined by the boundary conditions in \eqs{eq:cjump2} and (\ref{eq:normalj2}) at the 
interface, which lead to
  \begin{equation}
    Q' = \frac{D_1 - \lambda D_2}{2 D_+} Q ,
    \qquad
    Q'' = \frac{\lambda D_1}{D_+} Q ,
    \qquad
    D_+ := \frac{1}{2} \left(D_1 + \lambda D_2\right) .
  \end{equation}
In this manner, \eq{eq:cmonopole} in the main text is derived.

The Marangoni stress associated with the number density profile, given by 
\eq{eq:cmonopole} in the main text, is
\begin{equation}
  \label{eq:marangonistress}
  \nabla_\parallel \gamma(\br)
  = - b_0 \nabla_\parallel \left[ c(\br, z=0^+) - c_1^\infty \right]
  = \frac{Q b_0}{4\pi D_+} \frac{\br}{(r^2 + L^2)^{3/2}} .
\end{equation}
The Marangoni flow is obtained by evaluating \eq{eq:flowField} in the main text. To this 
end we introduce the Fourier transform of the Oseen tensor \cite{KiKa91}
\begin{subequations}
\begin{equation}
  \hat{\mathcal{O}}_{\bk, q} := \int d^2 \br \, dz \; \mathrm{e}^{-i\bk\cdot\br-i q z} 
  \mathcal{O} (\bx=\br+z\be_z)
  = \frac{1}{\eta_+ {(k^2+q^2)}}\left[\mathcal{I} 
    - {\frac{(\bk + q \be_z) ( \bk + q \be_z)}{k^2+q^2}} \right] ,
\end{equation}
as well as the Fourier transform of \eq{eq:marangonistress}:
\begin{equation}
  \int d^2 \br \; \mathrm{e}^{-i\bk\cdot\br} \nabla_\parallel \gamma(\br)
  = - i\bk \hat{g}_k ,
\end{equation}
\begin{equation}
  \label{eq:bgamma}
  { \hat{g}_k := b_0 \int d^2 \br \; \mathrm{e}^{-i\bk\cdot\br} 
    \left[ c(\br,z=0) -c_1^\infty \right] }
  = \frac{Q b_0}{4\pi D_+} \int d^2 \br \; 
  \frac{\mathrm{e}^{-i\bk\cdot\br}}{\sqrt{r^2+L^2}}
  = \frac{Q b_0}{2 D_+ k} \mathrm{e}^{-k L} \,.
\end{equation}
\end{subequations}
With this, the 2D convolution in \eq{eq:marangoni} can be expressed as
\begin{eqnarray}
  \label{eq:marangoniSI}
  \bu(\bx) & = & \int \frac{d^2\bk}{(2\pi)^2} \; \mathrm{e}^{i\bk\cdot\br}
  \; \hat{g}_k \, (- i\bk) \cdot 
  \int_{-\infty}^{+\infty} \frac{dq}{2\pi} \; \frac{\mathrm{e}^{iqz}}{\eta_+ (k^2+q^2)}
  \left[ \mathcal{I} - \frac{(\bk+q\be_z)(\bk+q\be_z)}{k^2+q^2} \right] 
  \nonumber \\
  & = & \frac{1}{16 \pi^2 \eta_+} \int d^2 \bk \; \mathrm{e}^{i\bk\cdot\br-k|z|}
   \; \hat{g}_k \frac{(- i\bk)}{k}\cdot \left[
    2 \mathcal{I} - \frac{\bk\bk}{k^2} \left(1 + k|z|\right) 
    - \be_z\be_z \left(1 - k|z|\right)
    - \left( \frac{\bk}{k} \be_z + \be_z \frac{\bk}{k} \right) i k z 
  \right] \nonumber \\
  & = & \frac{1}{16 \pi^2 \eta_+} \int d^2 \bk \; \mathrm{e}^{i\bk\cdot\br-k|z|}
  \hat{g}_k \left[
    \frac{i\bk}{k} \left( k|z| - 1 \right) - k z \be_z
  \right] \nonumber \\
  & = & \frac{1}{8\pi\eta_+} \int_0^{+\infty} dk\; k \mathrm{e}^{-k|z|} \hat{g}_k 
  \left[ \frac{\br}{r} ( 1 - k|z| ) J_1(k r) - \be_z k z J_0 (k r) \right] ,
\end{eqnarray}
by computing the integral over $q$, taking into account that $\be_z\cdot\bk=0$, 
and exploiting the radial symmetry of the configuration {which facilitates} the 
angular integration. With \eq{eq:bgamma} for $\hat{g}_k$, carrying out the last 
integral in \eq{eq:marangoniSI} renders \eq{eq:uradsym2} in the main text. 

The Marangoni flow field $\bu(\bx)$ given by \eq{eq:uradsym2} is, as 
expected, continuous and bounded even at the interface ($z = 0$):
\begin{equation}
  \bu(\br, z=0) = \be_r \frac{Q b_0}{16\pi D_+ \eta_+ r} 
    \left[ 1 - 
      \frac{L (r^2 + L^2)}{[ r^2 + L^2 ]^{3/2}}
    \right] 
    \sim \left\{
      \begin{array}[c]{cl}
        {\be_r} /r, & r\to\infty , \\
        & \\
        r{\be_r} , & r\to 0 .
     \end{array}
    \right.
\end{equation}
For $|\bx|=\sqrt{r^2 + z^2}\to \infty$ \eq{eq:marangoniSI} yields the
asymptotic behavior $|\bu{(\br,z)}|\propto 1/|\bx|$, which, however,
\emph{does not} correspond to a hydrodynamic monopole because this
asymptotic decay is not spherically symmetric.  Instead, $\bu(\bx)$ is
the solution of an inhomogeneous Stokes equation with a spatially
extended source (see \eq{eq:stokesSI}). The origin of the slow decay
$\sim 1/x$ of the field $\bu$ can be traced back to the equally slow
decay $\sim 1/r$ of the spatially varying surface tension (see
\eq{eq:uradsym2} in the main text), so that the capillary forces (in
the present case acting as Marangoni stresses) can be balanced by
hydrodynamic stresses, as stated by \eq{eq:tangentstress} in the main
text.


\section{Exact solution for the velocity $\boldsymbol{\cal V}$}

Since the system under consideration is axisymmetric and involves boundary conditions 
only at a spherical and at a planar surface, the diffusion and Stokes equations can be 
solved by employing bipolar coordinates. The bipolar coordinates $(\xi,\eta)$,
$-\infty<\xi<\infty$, $0\leq \eta \leq \pi$, are defined\footnote{We assume that the 
context suffices to disentangle the same notation $\eta$ for the bipolar coordinate 
$\eta$ and for the viscosity of the fluids.} such that the vertical coordinate $z$ and 
the radial distance $r$ from the $z$--axis are given by
\cite{HaBr73,Bren61}
\begin{equation}
  \label{eq:bipolar}
  z = \varkappa \frac{\sinh\xi}{\cosh\xi - \cos\eta} ,
  \qquad
  r = \varkappa \frac{\sin\eta}{\cosh\xi - \cos\eta} ,
\end{equation}
where $\varkappa =  R \sinh \xi_0$ and $\xi_0 = \mathrm{arccosh} (L/R) $ are chosen such 
that the {manifold} $\xi = \xi_0$ corresponds to the spherical surface of radius 
$R$ centered at $z = L$ ({which is} the surface of the particle). The plane {$z=0$} 
of the interface corresponds to $\xi = 0$. {In order to simplify the} notation we 
introduce $\omega := \cos \eta$. 

\subsection{Solution of the diffusion equation}
\label{sec:cbipolar}

\begin{figure}[!bhtb]
\includegraphics[width=\textwidth]{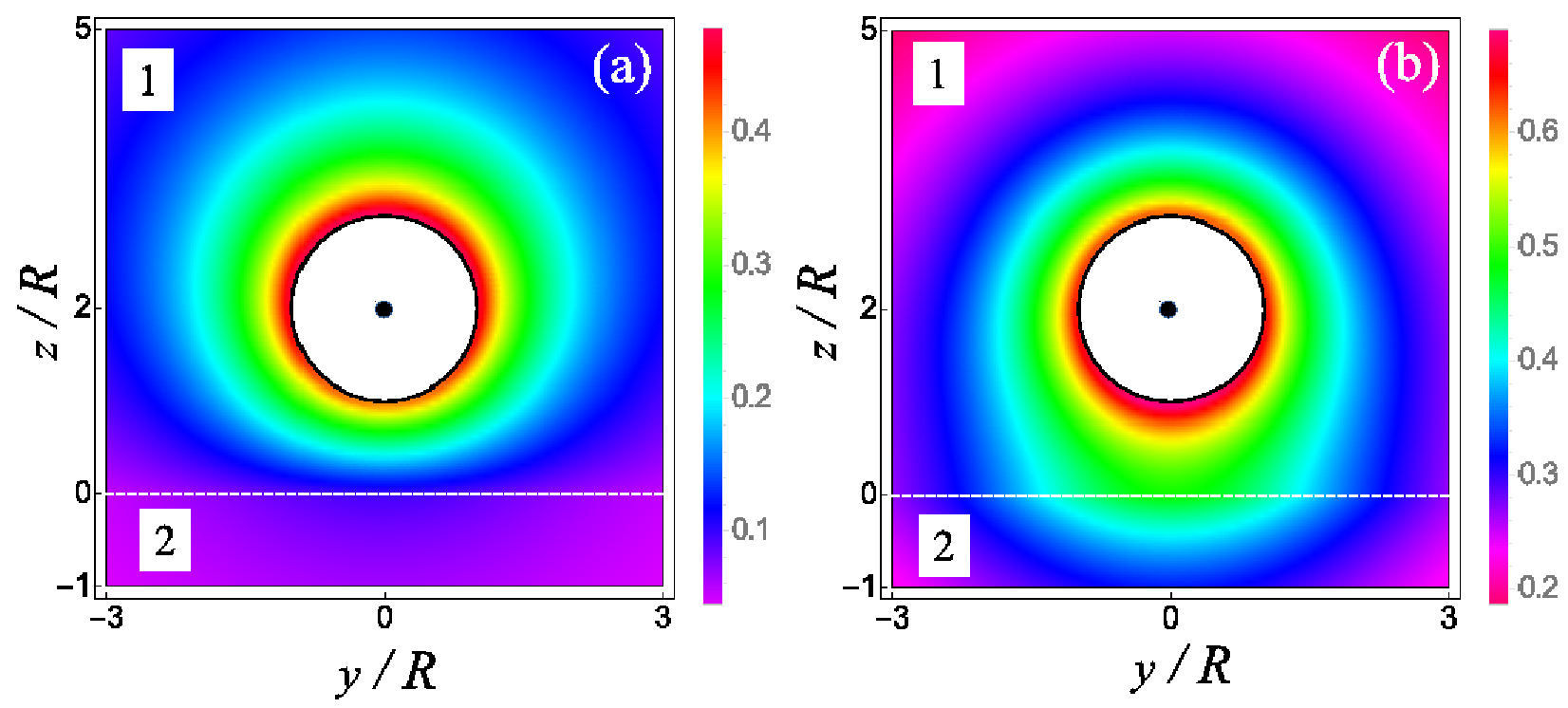}
\caption{\label{fig_dens_bipolar} Plots of the number density $c(\bx) - c_1^\infty$ 
(in units of $Q \sinh \xi_0/(4 \pi D_1 R)$, see \eq{eq:cbipolar2}) in the $y$-$z$ 
plane for a spherical particle located at $z/R = L/R = 2$ for $\lambda=1$ 
(i.e., $c_2^\infty = c_1^\infty$) and (a) $D_2/D_1 = 5$ and (b) $D_2/D_1 = 1/5$, 
respectively. Note that, for reasons of clarity concerning the details close to the 
interface $z=0$, the color coding scales in (a) and (b) are different. The whole surface 
of the particle is chemically active and homogeneous.}
\end{figure}

The axisymmetric solution to \eq{eq:diff} can be expressed in terms of the Legendre 
polynomials
$P_n$ as \cite{Jeffery1912}
\begin{equation}
  \label{eq:cbipolar}
  c(\bx) = {\cal C} + \frac{Q \sinh \xi_0}{4 \pi D_1 R} (\cosh\xi - \omega)^{1/2} 
\sum_{n=0}^{+\infty}
{  \left\lbrace 
    A_n \sinh \left[\left(n+\frac{1}{2}\right)\xi \right] + B_n 
\cosh \left[\left(n+\frac{1}{2}\right) \xi \right] \right \rbrace 
}
P_n(\omega)\,,~~\xi > 0\,,
\end{equation}
in fluid 1, with a similar expression but with different coefficients $\hat{\cal C}, 
\hat{A}_n$, and $\hat{B}_n$ in fluid 2 ($\xi < 0$). The prefactor $Q/(D_1 R)$ has the 
units of a number density, so that the coefficients $A_n, B_n, \hat{A}_n$, and $\hat{B}_n$
are dimensionless. (We note that the same prefactor $Q/(D_1 R)$ is used for both $\xi > 
0$ and $\xi < 0$; see also \eq{eq:Q''}.) These coefficients are determined by the 
boundary conditions expressed in Eqs.~(\ref{eq:mu02} -- \ref{eq:Qflux2}):
\begin{enumerate}

\item Since $|\bx| \to \infty$ maps {onto} $\{\xi = 0,\, \omega = 1\}$
  and because by construction the second term of \eq{eq:cbipolar}
  vanishes in this limit (irrespective of whether $\xi = {0^+}$ or
  $\xi = {0^-}$), the boundary condition~(\ref{eq:mu02}) 
    implies ${\cal C} = c_1^\infty$, $\hat{\cal C} = c_2^\infty$.

\item Since $\xi \to - \infty$ maps onto $\{r = 0,\, z =
  -\varkappa\}$, which is a point within fluid 2, the requirement
  that the density is bounded everywhere imposes that terms of the
  form $\exp[-(n+1/2) \xi]$ cannot occur in the series expansion
  corresponding to $\xi < 0$. This leads to $\hat{A}_n = \hat{B}_n$.
  There is no such requirement for $\xi\to +\infty$, because this
  maps onto the point $\{r = 0, z = \varkappa\}$ which is located
  inside the particle.

\item The boundary condition in \eq{eq:cjump2} concerning the
    discontinuity of $c(\bx)$ at the interface ($\xi = 0$) implies
    (recall that $\hat{\cal C}/{\cal C} = c_2^\infty/c_1^\infty=\lambda$) the relation 
$\lambda B_n = \hat{B}_n$ ($ = \hat{A}_n$, see above).

\item At the interface ($\xi = 0$) one has
  \begin{equation}
    \label{eq:normalderivative}
    \left. \frac{\partial c}{\partial z}\right|_{z=0^{\pm}} 
    = \left. \be_z\cdot\nabla c\right|_{z=0^{\pm}} 
    = \frac{1}{h_\xi} \left. \frac{\partial c}{\partial \xi}\right|_{\xi=0^{\pm}} 
  \end{equation}
  in terms of the corresponding scale factor $h_\xi =
  \dfrac{\varkappa}{\cosh \xi - \omega}$. Thus the continuity of the
  current along the direction of the interface normal (see
    \eq{eq:normalj2}) renders the condition
  \begin{equation}
    \label{eq:normalcurrent}
    \be_z\cdot \left[ D_1 \left. \nabla c \right|_{z=0^+} - 
      D_2 \left. \nabla c \right|_{z=0^-} 
    \right] = 0 
    \qquad\Rightarrow\qquad
    D_1 \left. \frac{\partial c}{\partial \xi}\right|_{\xi=0^+} 
    - D_2 \left. \frac{\partial c}{\partial \xi}\right|_{\xi=0^-} 
    = 0 .
  \end{equation}
  The series representation in \eq{eq:cbipolar} gives
  \begin{equation}
     \left. \frac{\partial c}{\partial \xi}\right|_{\xi=0^+} 
    = \frac{Q \sinh \xi_0}{4 \pi D_1 R} {(1 - \omega)^{1/2}}
    \sum_{n=0}^{+\infty} \left(n+\frac{1}{2}\right) A_n P_n(\omega) ,
  \end{equation}
  in fluid 1 ($z\to 0^+$) and a similar expression with $A_n$
  replaced by $\hat{A}_n$ in fluid 2 ($z\to 0^-$).
  Therefore, \eq{eq:normalcurrent} leads to $D_1 A_n = D_2 \hat{A}_n$.
  Since $\hat{A}_n = \lambda B_n$, it follows that $D_1 A_n =
    \lambda D_2 B_n$ and \eq{eq:cbipolar} turns into
\begin{subequations} 
\begin{equation}
  \label{eq:cbipolar2}
  c(\bx) = c_1^\infty + \frac{Q \sinh \xi_0}{4 \pi D_1 R} (\cosh\xi - \omega)^{1/2} 
  \sum_{n=0}^{+\infty}
  B_n 
  \left\lbrace
    \dfrac{\lambda D_2}{D_1}  \sinh \left[\left(n+\frac{1}{2}\right)\xi\right] + 
    \cosh \left[\left(n+\frac{1}{2}\right)\xi \right] \right\rbrace P_n(\omega)\,,\; 
  \xi > 0\,,
\end{equation}
in fluid 1 and, due to $\hat{A}_n = \hat{B}_n = \lambda B_n$ and $c_2^\infty = 
\lambda c_1^\infty$, into
\begin{equation}
  c(\bx) = \lambda c_1^\infty + \frac{\lambda Q \sinh \xi_0}{4 \pi D_1 R} 
  (\cosh\xi - \omega)^{1/2} \sum_{n=0}^{+\infty}
  B_n 
  \left\lbrace
    \sinh \left[\left(n+\frac{1}{2}\right)\xi\right] + 
    \cosh \left[\left(n+\frac{1}{2}\right)\xi \right] \right\rbrace P_n(\omega)\,,\; 
  \xi < 0\,,
\end{equation}
in fluid 2.
\end{subequations}

\item The dimensionless coefficients $B_n$ are determined by the boundary condition at 
the surface of the particle (see \eq{eq:cflux} in the main text). Since at 
$\mathcal{S}_p$ (i.e., at the surface $\xi=\xi_0$) the normal is $\bn = - \be_\xi$, one 
has
\begin{equation}
\label{eq:norm_deriv_conc}
  \bn\cdot\nabla c(\bx\in\mathcal{S}_p) 
  = - \frac{1}{h_\xi}
  \left.\frac{\partial c}{\partial \xi}\right|_{\xi=\xi_0} .
\end{equation}
Inserting \eq{eq:cbipolar2} into \eq{eq:norm_deriv_conc} and by using the recursion 
relation
\begin{equation}
 (2 n + 1) \omega P_n(\omega) = (n+1) P_{n+1} (\omega) + n P_{n-1} (\omega)\,,
 \quad n \geq 0\,,
\end{equation}
the orthogonality of the Legendre polynomials,
\begin{equation}
  \label{Leg_orto}
  \int_{-1}^{+1} ds\; P_n(s) P_m(s) = \frac{2}{2n+1} \delta_{n,m} \,, 
  \quad 
  n,m \geq 0\,,
\end{equation}
and the following expressions for the integral \cite{Jeffery1912}
\begin{equation}
  \int\limits_{-1}^{+1} d\omega \dfrac{P_n(\omega)}{(\cosh\xi_0 - \omega)^{1/2}} = 
  \dfrac{2 \sqrt{2}}{2 n +1} e^{-(n+1/2) \xi_0}\,, 
  \quad
  \xi_0 > 0\,,
\end{equation}
one can transform the boundary condition in \eq{eq:cflux} into a set of linear 
equations for the coefficients $B_n$:
\begin{eqnarray}
 \label{Bn_eqs}
 2\sqrt{2} \,e^{-(n+1/2) \xi_0} & = & 
 (n+1) (B_n- B_{n+1}) \left\{
   \dfrac{\lambda D_2}{D_1} \cosh\left[\left(n + \dfrac{3}{2} \right) \xi \right] 
   + 
   \sinh \left[ \left(n + \dfrac{3}{2} \right)\xi \right] 
 \right\}\nonumber\\
 & & \mbox{} + n (B_n- B_{n-1}) \left\{
   \dfrac{\lambda D_2}{D_1} \cosh\left[\left(n - \dfrac{1}{2} \right) \xi \right] 
   + 
   \sinh \left[ \left(n - \dfrac{1}{2} \right)\xi \right] 
 \right\} ,
 \qquad
 n \geq 0\,,
\end{eqnarray}
with the convention that $B_{-1} = 0$. 
\end{enumerate}
This infinitely large system of linear equations is truncated at a sufficiently large 
index $n=N$ and it is then solved numerically. In practice, the truncation, as well as the 
series representation, are converging very fast. We have found that $N = 50$ is sufficient 
for providing accurate results. As an illustrative example of a calculation following the 
steps outlined above, in Fig. \ref{fig_dens_bipolar} we show plots of the field $c(\bx)$.

\subsection{Solution of the  auxiliary problem in terms of bipolar coordinates}
\label{sec:psiaux}

Our auxiliary problem of a (passive) sphere approaching a flat and not
deforming fluid-fluid interface (\eq{eq:aux}) has been solved
previously in terms of bipolar coordinates by Bart \cite{Bart68} and
by Lee and Leal \cite{LeLe80}. Nevertheless, we provide here the
details of the solution for the auxiliary problem because the former
reference \cite{Bart68} provides only the sum of the coefficients in
the series representation of the solution (see below), which is
insufficient for our problem (see, c.f., \eq{eq:V_in_bipol}), while
the latter reference \cite{LeLe80} provides expressions for these
coefficients (Eqs.~(61)--(66) in Ref.~\cite{LeLe80}) which, however,
in certain limits do not reduce to those corresponding to
  the limiting cases of a solid--liquid interface ($\eta_2/\eta_1 \to \infty$) or of a
``free'', i.e., liquid--gas interface ($\eta_2/\eta_1 \to 0$) as derived by Brenner
\cite{Bren61}\footnote{For example, in both limits Eqs.~(64)--(66) in 
Ref.~\cite{LeLe80} lead to expressions which
    differ by a factor of $1/(2n + 1)$ within the sums from the
    corresponding Eqs.~(2.19) and (3.13) in Ref.~\cite{Bren61}.
    However, the results for the so-called Stokes' law correction shown in
    Fig. 2 in Ref.~\cite{LeLe80} seem to be in agreement with the
    behavior expected from Ref.~\cite{Bren61}. We therefore surmise that
    (some of) the expressions in Eqs.~(61)--(66) in Ref.~\cite{LeLe80}
    contain errors, which are of typographical nature (only).}.
The axial symmetry of the problem allows one to express the solution of the Stokes 
equations in terms of a dimensionless stream function $\psi_\mathrm{aux}(\bx)$ (see 
\eq{eq:stream}), which can be represented in bipolar 
coordinates \cite{Jeffery1912,Bren61}:
\begin{eqnarray}
\label{stream_func}
  \psi_\mathrm{aux}(\bx) = \frac{1}{(\cosh\xi-\omega)^{3/2}} \sum_{n=1}^{+\infty}
  & & \left\lbrace
    K_n {\cosh \left[\left(n-\frac{1}{2}\right)\xi\right]}
    + L_n {\sinh \left[\left(n-\frac{1}{2}\right)\xi\right]}
  \right. 
\nonumber \\
  & & \left. +  M_n {\cosh \left[\left(n+\frac{3}{2}\right)\xi\right]}
    + N_n {\sinh \left[\left(n+\frac{3}{2}\right)\xi\right]}
  \right\rbrace 
\times  {\cal G}_{n+1}^{-1/2}(\omega) \, ,\;~\xi > 0 \,;
\end{eqnarray}
a similar expression, but with different coefficients ${\hat K}_n$, ${\hat L}_n$, ${\hat 
M}_n$, and ${\hat N}_n$, holds for $\xi < 0$, where 
\begin{equation}
 {\cal G}_{n}^{-1/2}(\omega) = \dfrac{P_{n-2}(\omega) - P_{n}(\omega)}{2 n - 1}
\end{equation}
denotes the Gegenbauer polynomial of order $n$ and degree $-1/2$ {\cite{Bren61}}. The 
dimensionless coefficients $K_n$, $L_n$, $M_n$, and $N_n$, as well as the hatted 
ones, depend on $\xi_0$ and are determined by the boundary conditions {for} the velocity 
field (see \eq{eq:aux}). We proceed along the lines of the previous section for 
determining the stream function:

\begin{enumerate}
\item The stream function given by \eq{stream_func} automatically
  satisfies the boundary condition of vanishing velocity at infinity
  \cite{HaBr73}.

\item Boundedness of the fields everywhere in the fluid domain, in particular at 
the point $\xi \to -\infty$ located in fluid 2, imposes that none of the 
terms $\sim \exp[-(n-1/2) \xi]$ and $\sim \exp[-(n+3/2) \xi]$ is present in the 
series expansion corresponding to $\xi < 0$. This leads to
\begin{equation}
\label{finite_stream}
{\hat K}_n = {\hat L}_n\,, \qquad
{\hat M}_n = {\hat N}_n\,,
\qquad n \geq 1\,.
\end{equation}

\item Due to \eq{eq:stream}, the requirement that the normal velocity vanishes 
at the interface in both fluid domains translates into
\begin{equation}
  \left.\frac{\partial \psi_\mathrm{aux}}{\partial r}\right|_{z=0^\pm} = 0\,.
\end{equation}
Since the interface ($z=0$) coincides with the surface $\xi=0$, this means (see also 
Eq.~(3.3) in Ref.~\cite{Bren61})
\begin{equation}
  \left.\frac{\partial \psi_\mathrm{aux}}{\partial \omega}\right|_{\xi=0^\pm} = 0\,.
\end{equation}
This leads to
\begin{equation}
\label{normal_V_interf}
K_n = -M_n\,, \qquad
{\hat K}_n = -{\hat M}_n\,,
\qquad
n \geq 1\,{,}
\end{equation}
which implies necessarily the continuity of $\psi_\mathrm{aux}$ at the 
interface: $\psi_\mathrm{aux}(\xi = 0^+,\omega) = \psi_\mathrm{aux}(\xi = 
0^-,\omega) = 0$. The last relation together with \eq{finite_stream} implies
\begin{equation}
\label{hat_coeff_rel}
{\hat K}_n = {\hat L}_n = -{\hat M}_n =-{\hat N}_n\,, \qquad n \geq 1\,. 
\end{equation}

\item Due to \eq{eq:stream} and to an argument as the one used in
  \eq{eq:normalderivative} (see also Eq.~(8) in Ref.~\cite{Bart68}),
  continuity of the tangential velocity at the interface translates
  into
\begin{equation}
\label{eq:cont_der_stream_aux}
  \left.\frac{\partial \psi_\mathrm{aux}}{\partial z}\right|_{z=0^+}
  - \left.\frac{\partial \psi_\mathrm{aux}}{\partial z}\right|_{z=0^-} = 0 
  \qquad\Rightarrow\qquad
  \left.\frac{\partial \psi_\mathrm{aux}}{\partial \xi}\right|_{\xi=0^+} 
  - \left.\frac{\partial \psi_\mathrm{aux}}{\partial \xi}\right|_{\xi=0^-} = 0 ,
\end{equation}
from which one obtains
\begin{equation}
 \label{tang_V_interf}
\left(n-\dfrac{1}{2}\right) L_n +  \left(n + \dfrac{3}{2}\right) N_n = 
\left(n-\dfrac{1}{2}\right) {\hat L}_n +  \left(n + \dfrac{3}{2}\right) {\hat N}_n\,,
\qquad n \geq 1\,.
\end{equation}
Together with \eq{hat_coeff_rel}, this leads to
\begin{equation}
\label{hat_coeff}
{\hat K}_n = - \dfrac{1}{2} \left[\left(n-\dfrac{1}{2}\right) L_n +  
\left(n + \dfrac{3}{2}\right) N_n \right]\,,
\qquad n \geq 1\,.
\end{equation}

\item The last one of the boundary conditions at the interface is the
  continuity of the tangential stresses. (We remind the reader that
  the auxiliary problem involves an \textit{inert (i.e., non-active)} particle. 
  In this case there are no lateral variations of the surface tension of the 
interface. Therefore in the auxiliary problem there is no Maragoni stress induced, and
  thus \eq{eq:tangentstress} implies continuity of the tangential
  stresses.)  Using \eq{eq:stream} and the vanishing of the normal
  velocity at the interface reduces \eq{eq:tangentstress} in the main
  text to (see also Eqs.~(3.1)--(3.4) in Ref.~\cite{Bren61} for
  expressing the stress tensor components in terms of the stream
  function)
  \begin{equation}
  \label{eq:stress_cont_via_stream}
    \eta_1 \left.\frac{\partial^2\psi_\mathrm{aux}}{\partial z^2}\right|_{z=0^+} 
    - \eta_2 \left.\frac{\partial^2\psi_\mathrm{aux}}{\partial z^2}\right|_{z=0^-} 
    = 0 
    \qquad\Rightarrow\qquad
    \eta_1 \left.\frac{\partial^2\psi_\mathrm{aux}}{\partial \xi^2}\right|_{\xi=0^+} 
    - \eta_2 \left.\frac{\partial^2\psi_\mathrm{aux}}{\partial \xi^2}\right|_{\xi=0^-} 
    = 0 .
  \end{equation}
  The latter equation follows from iterating an argument as the one given in 
\eq{eq:normalderivative} (see also Eqs.~(3.1)--(3.4)
  in Ref.~\cite{Bren61}). By taking advantage of the continuity of
  $\psi_\mathrm{aux}$ at $\xi=0$, which facilitates the simplification of the 
expressions for the second derivatives evaluated at the interface\footnote{With the
    notations (the prime denotes the derivative with respect to
    $\xi$) $f(\xi,\omega) = (\cosh \xi - \omega)^{-3/2}$,
    $g(\xi,\omega) = f'(\xi,\omega)/f(\xi,\omega)$,
    and $\Phi(\xi,\omega) =
    \psi_\mathrm{aux}(\xi,\omega)/f(\xi,\omega)$, one has
    $\psi_\mathrm{aux}''=g' \psi_\mathrm{aux}+g(\psi_\mathrm{aux}'+f
    \Phi')+f \Phi''$. Noting that at the interface ($\xi = 0$) both $g$ and 
$\psi_\mathrm{aux}$ vanish (the latter due to point 3 above), while $f$ is continuous 
across the interface, one infers that the jump of $\psi_\mathrm{aux}''$ at the 
interface implies that at the interface $\Phi''$ exhibits a jump $\psi_\mathrm{aux}''/f$, 
from which \eq{eq:tang_stress_interf} follows.}, one arrives at
\begin{equation}
\label{eq:tang_stress_interf}
\eta_1 \left[ \left(n-\dfrac{1}{2}\right)^2 K_n + 
\left(n + \dfrac{3}{2}\right)^2 M_n \right] 
= \eta_2 \left[ \left(n-\dfrac{1}{2}\right)^2 {\hat K}_n +  
\left(n + \dfrac{3}{2}\right)^2 
{\hat M}_n \right]\,.
\end{equation}
By combining {\eq{eq:tang_stress_interf}} with \eqs{normal_V_interf} and 
(\ref{hat_coeff}) one obtains 
\begin{equation}
 \label{nonhat_hat_coef}
K_n = \frac{\eta_2}{\eta_1} {\hat K}_n = -  \dfrac{\eta_2}{2\eta_1} 
\left[\left(n-\dfrac{1}{2}\right) L_n + \left(n + \dfrac{3}{2}\right) N_n \right] 
\,, 
\qquad n \geq 1\,.
\end{equation}

\item Turning now to the boundary conditions at the sphere, the no-slip requirement 
(\eq{eq:V}) implies that at $\mathcal{S}_p$ $\bv_\mathrm{aux}$ has only a 
$z$ component,  which is equal to ${\cal V}_\mathrm{aux}$. Using the representation of 
$\bv_\mathrm{aux}$ in terms of the stream function (\eq{eq:stream}) one obtains
    \begin{equation}
    \label{vel_on_sph_stream}
      \left.\frac{\partial \psi_\mathrm{aux}}{\partial z} \right|_{\bx\in\mathcal{S}_p} 
      = 0 ,
      \qquad
      \left.- \frac{R^2}{r} \frac{\partial \psi_\mathrm{aux}}{\partial r} 
\right|_{\bx\in\mathcal{S}_p} 
      = 1 .
    \end{equation}
Using \eq{vel_on_sph_stream} and the chain rule for computing nested derivatives 
(see also Eqs.~(2.3) and (2.4) in Ref.~\cite{Bren61}), one obtains the following 
equivalent conditions (the surface of the particle is given by $\xi=\xi_0$):
    \begin{equation}
    \label{eq:bipol_deriv_psi}
      \left.\frac{\partial \psi_\mathrm{aux}}{\partial \omega} \right|_{\xi=\xi_0} 
      = - \frac{\partial}{\partial \omega}\left( \frac{r^2}{2 R^2} \right)_{\xi=\xi_0} ,
      \qquad
      \left.\frac{\partial \psi_\mathrm{aux}}{\partial \xi} \right|_{\xi=\xi_0} 
      = - \frac{\partial}{\partial \xi}\left( \frac{r^2}{2 R^2} \right)_{\xi=\xi_0} .
    \end{equation}
The first of these two equations can be integrated over $\omega$. Since $\omega=+1$ 
implies $r=0$ (see \eq{eq:bipolar}) and $\psi_\textrm{aux}(\xi_0,\omega=+1)=0$ (see 
\eq{stream_func} with ${\cal G}_{n+1}^{-1/2}(1)=0$), one finds \cite{Bren61}
    \begin{subequations}
      \label{no_slip_surf}
      \begin{equation}
        \label{no_slip_surf_omega}
      {\left.{\psi_\mathrm{aux}}(\xi,\omega)\right|_{\xi=\xi_0} = - \left.
      \frac{r^2}{2 R^2} \right|_{\xi=\xi_0} \quad\Rightarrow\quad}
        \left. \left(\cosh \xi - \omega\right)^{3/2} \psi_\mathrm{aux}(\xi,\omega) 
        \right|_{\xi=\xi_0}
        = - \left. 
          \left(\cosh \xi - \omega\right)^{3/2} \frac{r^2}{2 R^2} \right|_{\xi=\xi_0} .
      \end{equation}
The second equation in \eq{eq:bipol_deriv_psi} can be cast into the form
      \begin{equation}
        \left.\frac{\partial}{\partial \xi} \left[ 
            \left(\cosh \xi - \omega\right)^{3/2} {\psi_\mathrm{aux}}(\xi,\omega) \right] 
        \right|_{\xi=\xi_0}
        = - \left. \frac{\partial}{\partial \xi}\left[
            \left(\cosh \xi - \omega\right)^{3/2} \frac{r^2}{2 R^2} \right]
        \right|_{\xi=\xi_0} .
      \end{equation}
This can be checked by differentiating the products on each side and by using 
\eq{no_slip_surf_omega}.
    \end{subequations}
By using the expansion (with $\varkappa = R \sinh \xi_0$ and $r = \varkappa 
\sqrt{1-\omega^2}/(\cosh \xi -\omega)$) \cite{Bren61}
\begin{equation}
(\cosh\xi-\omega)^{3/2}\, r^2 = \sqrt{2} \, \varkappa^2 \sum_{n = 1}^{\infty} n (n+1) 
\left[\dfrac{e^{-(n-1/2) \xi}}{2 n -1} - \dfrac{e^{-(n+3/2) \xi}}{2 n + 3} \right]
\times  {\cal G}_{n+1}^{-1/2}(\omega)
\end{equation}
and by employing the orthogonality of the Gegenbauer polynomials 
$\{{\cal G}_{n}^{-1/2}\}_{n \geq 2}$ one can transform \eq{no_slip_surf} into equalities 
of 
individual terms: 
\begin{subequations}
\label{no_slip_stream}
\begin{eqnarray}
K_n {\cosh \left[\left(n-\frac{1}{2}\right)\xi_0\right]}
& + & L_n {\sinh \left[\left(n-\frac{1}{2}\right)\xi_0\right]}
+ M_n {\cosh \left[\left(n+\frac{3}{2}\right)\xi_0\right]}
+ N_n {\sinh \left[\left(n+\frac{3}{2}\right)\xi_0 \right]}
\nonumber\\
& = & - \dfrac{\sqrt{2}}{4} (\sinh \xi_0)^2\,
n (n+1) \left[\dfrac{e^{-(n-1/2)\xi_0}}{n-1/2} - \dfrac{e^{-(n+3/2)\xi_0}}{n+3/2} \right]
\,, 
\qquad n \geq 1 \,,
\end{eqnarray}
and
\begin{eqnarray}
  &&\left(n-\frac{1}{2}\right) 
  \left\lbrace K_n \sinh
   \left[\left(n-\frac{1}{2}\right)\xi_0 \right]\right.
  + \left. L_n \cosh \left[\left(n-\frac{1}{2}\right)\xi_0\right] \right\rbrace
\nonumber\\
  &+& \left(n+\frac{3}{2}\right) 
  \left\lbrace M_n \sinh \left[\left(n+\frac{3}{2}\right)\xi_0\right]
    + N_n \cosh \left[\left(n+\frac{3}{2}\right)\xi_0\right] \right \rbrace
  \nonumber\\
  &=& - \dfrac{\sqrt{2}}{4} (\sinh \xi_0)^2\,
  n (n+1) \left[- e^{-(n-1/2)\xi_0} + e^{-(n+3/2)\xi_0} \right]\,, 
  \qquad n \geq 1 \,.
\end{eqnarray}
\end{subequations}
By using \eqs{normal_V_interf} and (\ref{nonhat_hat_coef}) in order to replace $K_n$ and 
$M_n$ in \eq{no_slip_stream}, for every $n \geq 1$ one obtains a system of linear 
equations with the unknowns $L_n$ and $N_n$, the solution of which is
\begin{subequations}
 \label{stream_coeff}
\begin{eqnarray}
L_n &=& - \dfrac{\sqrt{2}}{4} (\sinh \xi_0)^2 \, n (n+1) 
\dfrac{\chi_n^{(1)} \beta_n^{(2)} - \chi_n^{(2)} \beta_n^{(1)}}
{\alpha_n^{(1)} \beta_n^{(2)} - \alpha_n^{(2)} \beta_n^{(1)}}\,, \\
N_n &=& - \dfrac{\sqrt{2}}{4} (\sinh \xi_0)^2 \, n (n+1) 
\dfrac{\chi_n^{(2)} \alpha_n^{(1)} - \chi_n^{(1)} \alpha_n^{(2)}}
{\alpha_n^{(1)} \beta_n^{(2)} - \alpha_n^{(2)} \beta_n^{(1)}}\,,
\end{eqnarray}
\end{subequations}
where
\begin{eqnarray}
\chi_n^{(1)} &=& \dfrac{e^{-(n-1/2)\xi_0}}{n-1/2} - \dfrac{e^{-(n+3/2)\xi_0}}{n+3/2}
\,,\nonumber\\
\chi_n^{(2)} &=& - e^{-(n-1/2)\xi_0} + e^{-(n+3/2)\xi_0}\,,\nonumber\\
\alpha_n^{(1)} &=& {\sinh \left[\left(n-\frac{1}{2}\right)\xi_0\right] 
+ \dfrac{\eta_2}{2\eta_1} \left(n-\frac{1}{2}\right) 
\left\lbrace \cosh \left[\left(n+\frac{3}{2}\right)\xi_0\right] - \cosh \left[
\left(n-\frac{1}{2}\right)\xi_0\right] \right\rbrace}
\,,\nonumber\\
\beta_n^{(1)} &=& {\sinh \left[\left(n+\frac{3}{2}\right)\xi_0 \right]
+ \dfrac{\eta_2}{2\eta_1} \left(n+\frac{3}{2}\right) 
\left\lbrace \cosh \left[\left(n+\frac{3}{2}\right)\xi_0\right] - \cosh \left[
\left(n-\frac{1}{2}\right)\xi_0 \right]\right\rbrace}
\,, \\
\alpha_n^{(2)} &=& {\left(n-\frac{1}{2}\right) \left\lbrace 
\cosh \left[\left(n-\frac{1}{2}\right)\xi_0 \right] + \dfrac{\eta_2}{2\eta_1} 
\left\lbrace\left(n+\frac{3}{2}\right) 
\sinh \left[\left(n+\frac{3}{2}\right)\xi_0 \right] - \left(n-\frac{1}{2}\right) 
\sinh \left[\left(n-\frac{1}{2}\right)\xi_0 \right] \right\rbrace 
\right\rbrace}\,,\nonumber\\
\beta_n^{(2)} &=& {\left(n+\frac{3}{2}\right) \left\lbrace 
\cosh \left[\left(n+\frac{3}{2}\right)\xi_0\right] + \dfrac{\eta_2}{2\eta_1} 
\left\lbrace\left(n+\frac{3}{2}\right) 
\sinh \left[\left(n+\frac{3}{2}\right)\xi_0\right] - \left(n-\frac{1}{2}\right)
\sinh \left[\left(n-\frac{1}{2}\right)\xi_0 \right]\right\rbrace\right\rbrace} \,. 
\nonumber
\end{eqnarray}
\end{enumerate}
The coefficients $K_n$ and, due to \eq{normal_V_interf}, $M_n = - K_n$ and ${\hat K}_n$ 
are determined by \eqs{nonhat_hat_coef} and (\ref{hat_coeff}). The other hatted 
coefficients follow from \eqs{hat_coeff_rel}, (\ref{eq:tang_stress_interf}), and
(\ref{nonhat_hat_coef}). This concludes the derivation.

\begin{figure}[!t]
\includegraphics[width= \textwidth]{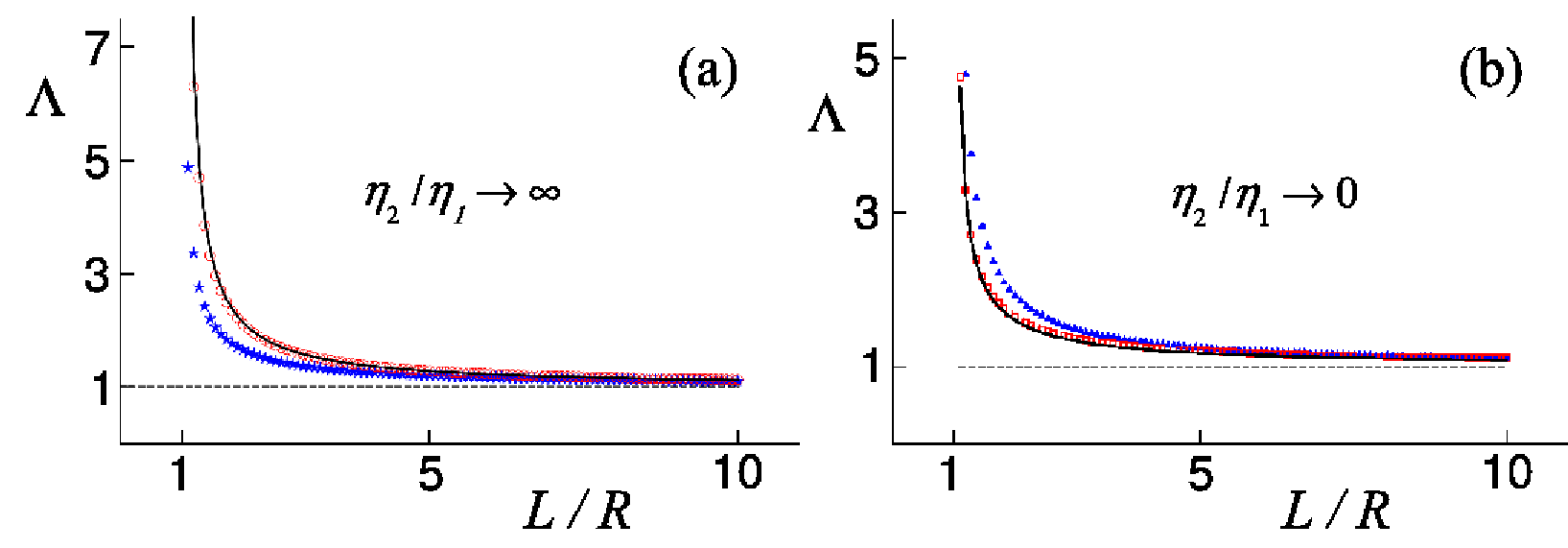}
\caption{\label{fig_comp_Bren} The ``Stokes' law correction''
  $\Lambda(L/R ; \eta_2/\eta_1)$ (\eq{Lam_def}) as a function of
  $L/R$ for (a) $\eta_2/\eta_1 = 0.1 (\tb{\star}),\, 20 (\tr{\circ})$ and (b)
  $\eta_2/\eta_1 = 2 (\tb{\textrm{\footnotesize $\blacktriangle$}}),\,
  0.05 (\tr{\textrm{\tiny $\square$}})$, respectively. The plots
  illustrate the convergence of $\Lambda(L/R ; \eta_2/\eta_1)$ in the
  limits $\eta_2/\eta_1 \to {\infty}$ (a) and $\eta_2/\eta_1 \to 0$
  (b), respectively, towards the limiting curves (solid lines)
  corresponding to (a) a particle approaching a solid wall
  {(}Eq.~(2.19) in Ref.~\cite{Bren61}{)} and (b) a particle
  approaching a ``free'' {(liquid--gas)} interface {(}Eq.~(3.13) in
  Ref.~\cite{Bren61}{)}. The dashed lines show the expected asymptotic
  behavior $\Lambda(L/R \to \infty ; \eta_2/\eta_1) = 1$.  }
\end{figure}

Once the coefficients $K_n, M_n, L_n$, and $N_n$ are known, one can
determine the force $\bF_\mathrm{aux}$ exerted by the particle on the
fluid (compare \eq{eq:faux}) from the stream function
$\Psi_\mathrm{aux}~{= \mathcal{V}_\mathrm{aux} R^2
  \psi_\mathrm{aux}(\bx)}$ as \cite{Jeffery1912,Bren61,Bart68,LeLe80}:
\begin{eqnarray}
\label{Lam_def}
\bF_\mathrm{aux} &=& - \be_z \dfrac{2 \sqrt{2} \pi \eta_1}{\varkappa} 
{\mathcal{V}}_\mathrm{aux} R^2 
\sum_{n=1}^\infty (K_n + L_n + M_n + N_n) 
= - \be_z \dfrac{2 \sqrt{2} \pi }{\sinh \xi_0} \eta_1 R {\mathcal{V}}_\mathrm{aux} 
\sum_{n=1}^\infty (L_n + N_n)
\nonumber\\
&=:& \be_z (6 \pi \eta_1 R  {\mathcal{V}}_\mathrm{aux})  \, \Lambda(L/R ; 
\eta_2/\eta_1)\,.
\end{eqnarray}
In the limits $\eta_2/\eta_1 \to \infty$ and $\eta_2/\eta_1 \to 0$, the so-called 
\cite{Bren61} ``Stokes' law correction'' $\Lambda(L/R ; \eta_2/\eta_1)$ 
should recover the exact results obtained by Brenner (Eqs.~(2.19) and (3.13) in
Ref.~\cite{Bren61}) for the case of a particle approaching a hard wall
(i.e., $\eta_2\to \infty$) and a ``free'' (liquid--gas) interface
(i.e., $\eta_2 = 0$), respectively. As shown in
Fig.~\ref{fig_comp_Bren}, $\Lambda(L/R ; \eta_2/\eta_1)$ indeed
exhibits this expected behavior, which provides a welcome check of our
derivation.

\subsection{Calculation of the drift velocity {$\boldsymbol{\cal V}$ in terms of} 
bipolar coordinates}

In this subsection we evaluate \eq{eq:faxen2} in the main text with the
  expressions derived for the fields $c(\bx)$ and $\psi_\mathrm{aux}(\bx)$. Since the 
interface ($z=0$) coincides with the surface $\xi=0$, in the integral in 
\eq{eq:faxen2} one has
  \begin{equation}
    dr\; \left.\frac{\partial c}{\partial r}\right|_{z = 0^+} 
    = d\omega \; \left.\frac{\partial c}{\partial \omega}\right|_{\xi = 0^+} \,.
  \end{equation}
Therefore, from \eq{eq:cbipolar2} one finds
  \begin{equation}
    dr\; \left.\frac{\partial c}{\partial r}\right|_{z=0^+} = 
    \frac{d\omega}{\sqrt{1-\omega}}\; 
    \frac{Q \sinh\xi_0}{4 \pi D_1 R} 
    \sum_{n=0}^{\infty} B_n \left[ 
      (1-\omega) \frac{dP_n}{d\omega} - \frac{1}{2} P_n(\omega)
    \right] .
  \end{equation}
Equation (\ref{stream_func}) leads to (see also \eqs{eq:normalderivative} and 
 (\ref{eq:cont_der_stream_aux}))
  \begin{equation}
    \left.\frac{\partial\psi_\mathrm{aux}}{\partial z}\right|_{z=0} 
    = \frac{1}{\varkappa \sqrt{1-\omega}}
    \sum_{n=1}^{\infty} \left[ 
      \left(n-\frac{1}{2}\right) L_n 
      + \left(n+\frac{3}{2}\right) N_n 
    \right] \times {\cal G}_{n+1}^{-1/2}(\omega),
  \end{equation}
  so that the integral in \eq{eq:faxen2} turns into
\begin{eqnarray}
  \int_0^\infty dr \; 
  \left.\frac{\partial c}{\partial r}\right|_{z=0^+}
  \left.\frac{\partial\psi_\mathrm{aux}}{\partial z}\right|_{z=0}
  & = & \frac{Q}{4 \pi D_1 R^2} \sum_{n=1}^{\infty} 
  \dfrac{1}{2 n + 1} \left[ 
    \left(n-\frac{1}{2}\right) L_n 
    + \left(n+\frac{3}{2}\right) N_n 
  \right] 
  \sum_{m=0}^{\infty} B_m
  \nonumber \\
  & \times &  \int_{-1}^{+1} d\omega\; \frac{1}{1-\omega} 
  \left[ P_{n-1}(\omega) - P_{n+1}(\omega) \right] 
  \left[ 
    (1-\omega) \frac{dP_m}{d\omega} - \frac{1}{2} P_m(\omega)
  \right]{\,,}  
\end{eqnarray}
with $B_m$ given by \eq{Bn_eqs}. From the identities (see Subsect.~\ref{sec:Pidentities})
\begin{subequations}
  \label{Leg_iden}
\begin{equation}
\label{Leg_iden_1}
  \int_{-1}^{+1} d\omega\; 
  \left[ P_{n-1}(\omega) - P_{n+1}(\omega) \right] \frac{d P_m}{d \omega} 
  = 2 \delta_{n,m} \left( 1 - \delta_{m,0} \right) ,
\end{equation}
and
\begin{equation}
\label{Leg_iden_2}
  \frac{1}{2} \int_{-1}^{+1} d\omega\; \frac{1}{1-\omega} 
  \left[ P_{n-1}(\omega) - P_{n+1}(\omega) \right] P_m(\omega)
  = \left\{
    \begin{array}[c]{cl}
      (2n+1)/[n(n+1)], & m < n , \\
      1/(n+1), & m=n , \\
      0 , & n < m ,
    \end{array}
  \right.
\end{equation}
\end{subequations}
which are valid for $n\geq 1$ and $m\geq 0$, one obtains for the sum over $m$ 
the simple expression
\begin{equation}
\label{m_sum} 
\sum_{m =0}^{\infty} B_m \times \int_{-1}^{+1} d\omega 
{\frac{1}{1-\omega} 
  \left[ P_{n-1}(\omega) - P_{n+1}(\omega) \right] 
  \left[ 
    (1-\omega) \frac{dP_m}{d\omega} - \frac{1}{2} P_m(\omega)
  \right]
}= \dfrac{2 n +1}{n+1} 
\left[B_n - \dfrac{1}{n} \sum_{m=0}^{n-1} B_m \right]\,,
\end{equation}
and thus finally arrives at
\begin{equation}
  \int_0^\infty dr \; 
  \left.\frac{\partial c}{\partial r}\right|_{z= 0^+}
  \left.\frac{\partial\psi_\mathrm{aux}}{\partial z}\right|_{z=0}
  = \frac{Q}{4 \pi D_1 R^2} \sum_{n=1}^\infty 
  \frac{1}{n+1} \left[ 
    \left(n-\frac{1}{2}\right) L_n 
    + \left(n+\frac{3}{2}\right) N_n 
  \right] 
  \left[ B_n - \frac{1}{n} \sum_{m=0}^{n-1} B_m 
  \right] .
\end{equation}
The force $\bF_\mathrm{aux}$ needed to drag the particle with velocity $\be_z 
\mathcal{V}_\mathrm{aux}$ is given by \eq{Lam_def}. Thus, the mobility $\Gamma_z$ for 
the auxiliary problem (see the text below \eq{eq:stream}) is given by
\begin{equation}
  \label{eq:gammaz}
 \Gamma_z := \dfrac{{\cal V}_\mathrm{aux}}{\be_z \cdot \bF_\mathrm{aux}} = - 
\frac{\sinh\xi_0}{2 \sqrt{2} \pi \eta_1 R} 
  \left[ \sum_{n=1}^\infty (L_n+N_n) \right]^{-1} .
\end{equation}
Therefore, {the expression for} the velocity {$\boldsymbol{\cal V}$} given by 
\eq{eq:faxen2} and normalized by $u(\bx_0 {= (0,0,L)})$ {(see Eqs.~(\ref{eq:uradsym2}) 
and (\ref{eq:VlargeL}) in the main text) can be written (using 
$\cosh\xi_0 = L/R$) as}
\begin{eqnarray}
\label{eq:V_in_bipol}
\dfrac{{\boldsymbol{\cal V}} \cdot \be_z}{\bu(\bx_0) \cdot \be_z} &=& -
\sqrt{2} 
  \left(1+\frac{\lambda D_2}{D_1}\right)
  \left(1+\frac{\eta_2}{\eta_1}\right) \,\sinh (2 \,\xi_0) 
   \left[ \sum_{n=1}^\infty (L_n+N_n) \right]^{-1}
  \nonumber \\
  & & \times
  \sum_{n=1}^\infty
  \frac{1}{n+1} \left[ 
    \left(n-\frac{1}{2}\right) L_n 
    + \left(n+\frac{3}{2}\right) N_n 
  \right] 
  \left[ B_n - \frac{1}{n} \sum_{m=0}^{n-1} B_m \right]\,{,}
\end{eqnarray}
which is the function plotted in Fig.~\ref{fig:V} of {the main text}.

\subsection{\label{sec:Pidentities} Derivation of \eq{Leg_iden}}

We note that due to $P_0 = 1$ the integral on the left hand side (lhs) of \eq{Leg_iden_1} 
is zero for $m = 0$. For $m \neq 0$ we employ the relation
\begin{equation}
 \dfrac{d}{d \omega} \left[ P_{n+1}(\omega) - P_{n-1}(\omega) \right] = (2n+1) 
P_n(\omega)\,,
\end{equation}
which can be obtained by straightforward differentiation of the defining 
expression for the Legendre polynomials (Rodriguez formula). This allows one to 
transform the lhs of \eq{Leg_iden_1} as follows:
\begin{eqnarray}
  \int_{-1}^{+1} d\omega\; 
  \left[P_{n-1}(\omega) - P_{n+1}(\omega) \right] \frac{d P_m}{d \omega} 
  &=& \left\lbrace P_m(\omega)\left[P_{n-1}(\omega) - P_{n+1}(\omega) 
\right]\right\rbrace_{{\omega = -1}}^{{\omega = +1}}
- \int_{-1}^{+1} d\omega\; P_m(\omega) 
\dfrac{d}{d \omega} \left[ P_{n-1}(\omega) - P_{n+1}(\omega) \right] \nonumber\\
 &=& (2n+1) \int_{-1}^{+1} d\omega\; P_m(\omega) P_n(\omega) = 2 \delta_{n,m}\,,
\end{eqnarray}
where in the first line the first term on the rhs vanishes due to 
$P_n(1) = 1,~ P_n(-1) = (-1)^n$, $n \geq 0$, and the last equality follows from the 
orthogonality of the Legendre polynomials (\eq{Leg_orto}). Combining the results 
for $m = 0$ and $m \neq 0$ by accounting for the vanishing of the integral for $m=0$ 
via the factor $(1 - \delta_{m,0})$ completes the derivation of \eq{Leg_iden_1}.

The identity
\begin{equation}
  P_{n-1}(\omega) - P_{n+1}(\omega) = \frac{2n+1}{n(n+1)} 
  (1-\omega^2) \frac{d P_n}{d\omega} ,
  \qquad
  n\geq 1,
\end{equation}
follows from the recursion relations for the Legendre polynomials.
Therefore, the integral $I_{m,n}$ in \eq{Leg_iden_2} can be written as
\begin{equation}
  \label{eq:Imn1}
  I_{m,n} := \frac{2n+1}{2n(n+1)} \int_{-1}^{+1} d\omega\; (1+\omega) P_m(\omega) 
  \frac{d P_n}{d\omega} ,
  \qquad
  m\geq 0, n\geq 1 .
\end{equation}
One also has the identities (summation theorems {\cite{GrRh07}}):
\begin{subequations}
  \begin{equation}
    \frac{dP_n}{d \omega} = 2 \sum_{k=0}^{[(n-1)/2]} \left(n - 2 k - \frac{1}{2} \right)
    P_{n-2k-1}(\omega) ,
    \qquad
    n\geq 1 ,
  \end{equation}
  and
  \begin{equation}
    \omega \frac{dP_n}{d \omega} = n P_n(\omega) 
    + 2 \left( 1 - \delta_{n,1} \right) 
    \sum_{k=0}^{[(n-2)/2]} \left(n - 2 k - \frac{3}{2} \right)
    P_{n-2k-2}(\omega) ,
    \qquad
    n\geq 1 ,
  \end{equation}
  where $[\cdots]$ denotes the integer part of the quantity in square brackets.
\end{subequations}
Inserting these expressions into \eq{eq:Imn1} and using the
orthogonality relations in \eq{Leg_orto} one finds
\begin{equation}
  \label{eq:Imn2}
  I_{m,n} = \frac{2n+1}{n(n+1)} \left[ \frac{n}{2n+1} \delta_{n,m} 
    + \sum_{k=0}^{[(n-1)/2]} \delta_{m,n - 2 k - 1}
    + \left( 1 - \delta_{n,1} \right) \sum_{k=0}^{[(n-2)/2]} \delta_{m,n - 2 k - 2} 
  \right] ,
  \qquad
  m\geq 0, n\geq 1 .
\end{equation}
In the sums, the corresponding indices take the values
\begin{subequations}
  \begin{equation}
    \sum_{k=0}^{[(n-1)/2]} \delta_{m,n - 2 k - 1} 
    = \left\{
      \begin{array}[c]{cl}
        \delta_{m,n - 1} + \delta_{m,n - 3} + \cdots + \delta_{m,0}, & \textrm{$n$ odd}, \\
        \delta_{m,n - 1} + \delta_{m,n - 3} + \cdots + \delta_{m,1}, & \textrm{$n$ even}, \\
      \end{array}
    \right.
  \end{equation}
  and
  \begin{equation}
    \sum_{k=0}^{[(n-2)/2]} \delta_{m,n - 2 k - 2} 
    = \left\{
      \begin{array}[c]{cl}
        \delta_{m,n - 2} + \delta_{m,n - 4} + \cdots + \delta_{m,1}, & \textrm{$n$ odd}, \\
        \delta_{m,n - 2} + \delta_{m,n - 4} + \cdots + \delta_{m,0}, & \textrm{$n$ even}, \\
      \end{array}
    \right.
  \end{equation}
\end{subequations}
so that \eq{eq:Imn2} can be written as
\begin{equation}
  I_{m,n} = \frac{2n+1}{n(n+1)} \left[ \frac{n}{2n+1} \delta_{n,m} 
    + \sum_{j=0}^{n-1} \delta_{m,j}
  \right] ,
  \qquad
  m\geq 0, n\geq 1 ,
\end{equation}
which leads to \eq{Leg_iden_2}.

\section{Particle drift induced by Marangoni flow}

In this section we estimate the particle drift due to the Marangoni flow generated by 
the chemical activity of the particle. In the limit $R/L \to 0$ one can use 
\eq{eq:VlargeL} in order to integrate the equation of motion
\begin{equation}
  \label{eq:drift}
  \frac{d L}{d t} = \be_z\cdot\boldsymbol{\mathcal{V}}
  \approx -\frac{Q b_0}{64\pi D_+ \eta_+ L}
\end{equation}
within the overdamped regime of the dynamics. Consistently with the physical
assumptions of the underlying model, this expression implies an ``adiabatic 
approximation'': the particle drift is sufficiently slow so that both the number 
density field $c(\bx)$ and the Marangoni flow can be computed as stationary solutions 
of the diffusion equation or of the Stokes equations, respectively. The solution of
\eq{eq:drift} with the arbitrary initial condition $L(0)=L_0$ is
\begin{equation}
  L^2(t) = L_0^2 - \frac{Q b_0}{32\pi D_+ \eta_+} t.
\end{equation}
Consider for simplicity the case $b_0>0$, $Q>0$, so that the the particle drifts towards 
the interface. We introduce the time $t_\mathrm{drift}$ by the condition 
$L(t_\mathrm{drift}) = R$, i.e., after $t_\mathrm{drift}$ the particle touches the 
interface. If initially $L_0\gg R$, the explicit solution above leads to
\begin{equation}
  \label{eq:tdrift}
  t_\mathrm{drift} = \frac{32\pi D_+ \eta_+ L_0^2}{Q b_0} .
\end{equation}
When compared with the diffusion time for the particle over the same
distance,
\begin{equation}
  \label{eq:tdiff}
  t_\mathrm{diff} = \frac{L_0^2}{D_p} 
  = \frac{6\pi\eta_+ R L_0^2}{k_B T},
\end{equation}
with the coefficient of diffusion of the colloidal particle $D_p$ as quoted in the main 
text, one obtains the ratio
\begin{equation}
  \label{eq:tratio}
  \frac{t_\mathrm{drift}}{t_\mathrm{diff}} = \frac{1}{2 q}
\end{equation}
in terms of the dimensionless parameter $q$ introduced in \eq{eq:q} in the main text. We 
point out that in the limit $L_0\gg R$ the drift time $t_\mathrm{drift}$ is 
independent of the specific value of the final separation $L(t_\mathrm{drift}) = R$. 
In particular, one could have as well set $R\to 0$ from the beginning so that $L 
\gg R$ is always fulfilled. This indicates that, during the drift, the particle spends 
most of the time in the spatial region where $L\gg R$, i.e., where the 
approximation for $\boldsymbol{\mathcal{V}}$, described by \eq{eq:VlargeL} in the main
text, is reliable. This provides a successful self--consistency check of the 
point--particle approximation employed in \eq{eq:drift}. Finally, in the opposite case 
($Q < 0$) that the particle drifts away from the interface, $t_\mathrm{drift}$ has to 
be interpreted as the time for the particle to reach a distance $\hat L \gg R$ with 
the initial condition $L(0) = R$, i.e., ${\hat L}^2 = R^2 - Q b_0 
t_\mathrm{drift}/(32\pi D_+ \eta_+)$. This notion of $t_\mathrm{drift}$ leads to 
\eq{eq:tratio}, too, but with $q$ replaced by $-q > 0$.

\section{\label{sec:approx} Consistency of the applied approximations}

Once a solution for the boundary-value problem formulated by
Eqs.~(\ref{eq:c}-\ref{eq:v}) is known, one can check \textit{a
  posteriori} the consistency of the underlying simplifying
approximations.

$\;$

\noindent {\textbullet}~First, our model assumes that the spatial distribution of the 
generated chemical species is approximately stationary and undistorted by the 
Marangoni flow. The relevance of the advection by this flow is quantified by the Peclet 
number for the number density field $c(\bx)$: perturbations on the relevant length
scale $L$ evolve by diffusion on a time scale\footnote{Here we use the 
``mean'' diffusion coefficient $D_+$ for the purpose of providing a simple estimate 
of the order of magnitude of this time scale.} $\propto L^2/D_+$, while the time 
scale associated with advection is $\propto L/|\bu(L)|=2 t_\mathrm{drift}(L)$ 
(see \eq{eq:VlargeL} in the main text and \eq{eq:tdrift}). This leads to a 
Peclet number $\mathrm{Pe}_\mathrm{chem}$ as the ratio of these two time 
scales. By using Eqs.~(\ref{eq:tdiff}) and (\ref{eq:tratio}) one obtains
\begin{equation}
  \label{eq:Pechem}
  \mathrm{Pe}_\mathrm{chem} 
    = \frac{L^2/D_+}{2 t_\mathrm{drift}}
    = |q| \frac{D_p}{D_+}.
\end{equation}
A small value of this ratio of time scales indicates that the assumption that $c(\bx)$ 
is a stationary field determined by diffusion (compare \eq{eq:c} in the main text) is 
well grounded. Typically one has $D_p \ll D_+$: a micron--sized particle 
exhibits $D_p\sim 10^{-13} \;\mathrm{m}^2/\mathrm{s}$, whereas $D_+ 
\sim 10^{-4}-10^{-9} \;\mathrm{m}^2/\mathrm{s}$ for the experimental set-ups 
discussed in the main text. Thus, even though the parameter $|q|$ can be large 
(see the discussion on the value of $|q|$ in the main text), $\mathrm{Pe}_\mathrm{chem}$ 
is small for colloidal particles. In the case that $D_p$ becomes comparable to 
$D_+$ (e.g., for nanoparticles and liquid-liquid interfaces) the range of 
values of $q$, for which the assumption of a diffusion--dominated stationary 
profile of $c(\bx)$ is expected to hold, is somewhat reduced.

$\;$

\noindent {\textbullet}~Another simplifying assumption is that of a flat fluid interface. 
We can derive an order-of-magnitude estimate of the actual interfacial deformation 
as follows. Upon integrating \eq{eq:stokesSI} over an infinitesimal cylindrical box which 
has the axis aligned with the interface normal $\be_z$, and by using Gauss' theorem, one 
obtains
  \begin{equation}
    \left[ 
      \left. \overset{\leftrightarrow}{\sigma}\right|_{z=0^+} - 
      \left.\overset{\leftrightarrow}{\sigma}\right|_{z=0^-}
    \right] \cdot \be_z 
    = - \nabla_\parallel\gamma - \be_z \phi .
  \end{equation}
From this with $\be_z \cdot  \nabla_\parallel\gamma = 0$ and the relationship 
$H=\phi/(2\gamma)$ for the constraining force $\phi$, it follows that the mean 
curvature $H$ is determined by discontinuity in the normal stress at the interface:
\begin{equation}
  \label{eq:H}
  H = - \frac{1}{2\gamma} \; \be_z \cdot
  \left[ 
      \left. \overset{\leftrightarrow}{\sigma}\right|_{z=0^+} - 
      \left.\overset{\leftrightarrow}{\sigma}\right|_{z=0^-}
    \right] \cdot \be_z .
\end{equation}
The velocity field $\bv(\bx)$ given by \eq{eq:fullflow} can be written as the sum 
$\bv(\bx) = \bu(\bx) + \bw(\bx)$ of the Marangoni flow $\bu(\bx)$ defined by 
\eq{eq:flowField} in the main text and a perturbation $\bw(\bx)$ thereof induced by the 
particle. The Marangoni flow can be shown to solve the following boundary--value 
problem \cite{JFD75}:
\begin{subequations}
  \label{eq:fullmarangoni}
  \begin{equation}
    \label{eq:sigmaMarangoni}
    \nabla\cdot\bu = 0 ,
    \qquad
    \nabla\cdot\overset{\leftrightarrow}{\sigma}_u 
    = - \delta(z) \nabla_\parallel \gamma ,
    \qquad
    \mathrm{if}\;\bx\in\mathcal{D} ,
  \end{equation}
  \begin{equation}
    \bu(\bx) \textrm{ continuous everywhere,}
  \end{equation}
  \begin{equation}
    \be_z\cdot\bu = 0 \textrm{ at the interface,}
  \end{equation}
  \begin{equation}
    \bu(|\bx|\to\infty)=0,
  \end{equation}
  \begin{equation}
    \overset{\leftrightarrow}{\sigma}_u(\bx) := \eta(\bx) \left[ 
      \nabla\bu + \left(\nabla\bu\right)^\dagger 
    \right] - p_u \mathcal{I} .
  \end{equation}
\end{subequations}
We note that in \eq{eq:sigmaMarangoni} no vertical force $\be_z\phi$ due to the 
Laplace pressure appears. Therefore, the Marangoni flow does not deform the interface 
and the constraint of a flat interface is actually exact.

The equations obeyed by the perturbation $\bw(\bx)$ are obtained by subtracting 
\eq{eq:fullmarangoni} from \eq{eq:fullflow}:
\begin{subequations}
  \label{eq:fullpert}
  \begin{equation}
    \nabla\cdot\bw = 0 ,
    \qquad
    \nabla\cdot\overset{\leftrightarrow}{\sigma}_w 
    = - \delta(z) \, \be_z \phi,
    \qquad
    \mathrm{if}\;\bx\in\mathcal{D} ,
  \end{equation}
  \begin{equation}
    \bw(\bx) \textrm{ continuous everywhere,}
  \end{equation}
  \begin{equation}
    \be_z\cdot\bw = 0 \textrm{ at the interface,}
  \end{equation}
  \begin{equation}
    \label{eq:Vw}
    \bw(\bx) = \bV - \bu(\bx) ,
    \qquad
    \mathrm{if}\;\bx\in\mathcal{S}_p ,
  \end{equation}
  \begin{equation}
    \bw(|\bx|\to\infty)=0,
  \end{equation}
  \begin{equation}
    \label{eq:sigmaw}
    \overset{\leftrightarrow}{\sigma}_w(\bx) := \eta(\bx) \left[ 
      \nabla\bw + \left(\nabla\bw\right)^\dagger 
    \right] - p_w \mathcal{I} .
  \end{equation}
\end{subequations}
Therefore, a possible interfacial deformation can be described completely in terms of 
the perturbation field $\bw(\bx)$ (compare \eq{eq:H}):
\begin{equation}
  \label{eq:Hw}
  H = - \frac{1}{2\gamma} \be_z \cdot
    \left[ 
      \left. \overset{\leftrightarrow}{\sigma}_w\right|_{z=0^+} - 
      \left.\overset{\leftrightarrow}{\sigma}_w\right|_{z=0^-}
    \right] \cdot \be_z .
\end{equation}
The field $\bw(\bx)$ is not identically zero only because the particle has a finite 
spatial extent: in the point--particle limit $R/L \to 0$, one obtains 
$\bV=\bu(\bx_0)$ (see \eq{eq:VlargeL} in the main text), and the boundary condition in 
\eq{eq:Vw} reduces to $\bw=0$. Therefore, the origin of the perturbation field is related 
to the variation of the Marangoni flow on the scale of the particle radius. In the 
limit $R/L\to 0$, the order of magnitude of the characteristic velocity scale of the 
perturbation can be estimated as (compare \eq{eq:uradsym2} in the main text)
\begin{equation}
  \label{eq:estimatew}
  |\bw| \sim |\bu(\bx\in\mathcal{S}_p)-\bu(\bx_0)| 
  ~\propto~ {R \left|\frac{\partial u_z}{\partial z}\right|_{\bx=\bx_0}}
  \propto~ \frac{R}{L} |\bu(\bx_0)|\,.
\end{equation}
This leads to the estimate
\begin{equation}
  \left|\frac{\partial w_z}{\partial z}\right|
  \sim \frac{|\bw|}{L}
  \sim \frac{R |\bu(\bx_0)|}{L^2} 
\end{equation}
because the scale of the spatial variation of $\bw$ is $L$ (compare 
\eq{eq:estimatew}). Assuming that this provides the correct order
of magnitude for the discontinuity at the interface in the normal stress 
corresponding to the perturbation field $\bw(\bx)$, one obtains from 
\eqs{eq:Hw} and  (\ref{eq:sigmaw}) the order of magnitude estimate\footnote{We use
    the arithmetic mean $\eta_+$ for the purpose of providing a simple
    estimate of the order of magnitude of the viscosity.}
\begin{equation}
   |H| \sim \frac{1}{2 \gamma} \left( \overset{\leftrightarrow}{\sigma}_w 
\right)_{zz} 
    \sim \frac{1}{2 \gamma} \eta_+ \left|\frac{\partial w_z}{\partial z}\right|
    ~\sim
    \frac{\eta_+ R |\bu(\bx_0)|}{2\gamma L^2}\,.
  \end{equation}
Using \eq{eq:VlargeL} for the value of $|\bu(\bx_0)|$ and the definition 
of $q$ in \eq{eq:q}, one finally arrives at
\begin{equation}
  L |H| \sim |q| \frac{k_B T}{12 \pi \gamma L^2} .
\end{equation}
The dimensionless product $L |H|$ quantifies the importance of the interfacial curvature 
on the relevant length scale $L$. A small value of $L|H|$ means that the approximation of 
a flat interface is reliable. With a typical value $\gamma \sim 10^7 k_B T / 
\mu\mathrm{m}^2$ of the surface tension one has
\begin{equation}
  L |H| \sim 3 \times 10^{-9} \times |q| \left(\frac{\mu\mathrm{m}}{L}\right)^2 .
\end{equation}
This is indeed a small quantity even if $|q|$ is as large as $10^4$ (see the discussion 
on the value of $|q|$ in the main text).


%